\documentclass[12pt]{article}
\usepackage{amsmath,amssymb,amsthm,amsxtra,bbm,overpic,bm,epsfig,subfigure,tikz,tikz-feynman}
\usepackage{hyperref}
\usepackage{mathrsfs}
\usepackage{enumitem}
\usepackage{graphicx}
\usepackage{color}
\usepackage{comment}
\usepackage{epstopdf}
\usepackage{float}
\usepackage{cite}
\allowdisplaybreaks[3]
\usetikzlibrary{arrows.meta}
\makeatletter
\tikzfeynmanset{compat=1.1.0} % 启用兼容模式（适用于 pdflatex）
\makeatother
\usepackage[all]{xy}

\usepackage{slashed,stmaryrd,multirow}
\numberwithin{equation}{section}

\def\thefootnote{\fnsymbol{footnote}}

\usepackage{hyperref}
\usepackage{slashed,stmaryrd,orcidlink}

\usepackage{lscape}%
\usepackage{array}
\usepackage{booktabs}%

\begin{document}
	
	\vspace{0.2cm}
	
	\begin{center}
		{\Large\bf One-loop Renormalization of the Type-I Seesaw Model \\in the On-shell Scheme}
	\end{center}
	
	\vspace{0.2cm}
	
	\begin{center}
		{\bf Jihong Huang}~{\orcidlink{0000-0002-5092-7002}},$^{1,2}$~\footnote{E-mail: huangjh@ihep.ac.cn}
		\quad
		{\bf Shun Zhou}~{\orcidlink{0000-0003-4572-9666}}~$^{1,2,3}$~\footnote{E-mail: zhoush@ihep.ac.cn (corresponding author)}
		\\
		\vspace{0.2cm}
		{\small
			$^{1}$Institute of High Energy Physics, Chinese Academy of Sciences, Beijing 100049, China\\
			$^{2}$School of Physical Sciences, University of Chinese Academy of Sciences, Beijing 100049, China\\
			$^{3}$Center for High Energy Physics, Henan Academy of Sciences, Zhengzhou 450046, China}
	\end{center}

	\vspace{0.5cm}
	
\begin{abstract}
In this paper, we continue to carry out the one-loop renormalization of the type-I seesaw model in the on-shell scheme. Different from the modified minimal-subtraction ($\overline{\rm MS}$) scheme, such an investigation is mainly motivated by the fact that the on-shell scheme has been widely adopted in the renormalization of the standard electroweak theory and implemented for its precision tests. We first specify the physical parameters in the on-shell scheme, and then fix the corresponding counterterms through on-shell renormalization conditions. In the presence of massive Majorana neutrinos, we propose a practical method to determine gauge-independent counterterms for the lepton flavor mixing matrix. With the explicit counterterms in both the $\overline{\rm MS}$ and on-shell schemes, we establish the matching relations of the electric charge, physical masses and flavor mixing matrix elements between these two schemes. Our results in the present and previous papers lay the foundation for precision calculations in the type-I seesaw model.
\end{abstract}

	\def\thefootnote{\arabic{footnote}}
	\setcounter{footnote}{0}
	
%	\newpage

	\newpage

\section{Introduction}
	
Extending the Standard Model (SM) with three right-handed neutrino singlets, the type-I seesaw model has been so far the simplest and most natural scenario to accommodate nonzero but tiny neutrino masses~\cite{Minkowski:1977sc,Yanagida:1979as,Gell-Mann:1979vob,Glashow:1979nm,Mohapatra:1979ia}. More explicitly, the generic gauge-invariant Lagrangian of the type-I seesaw model reads
\begin{eqnarray}
\label{eq:L_SS}
{\cal L}_{}^{} = {\cal L}_{\rm SM}^{} + \overline{N_{\rm R}^{}} {\rm i} \slashed{\partial} N_{\rm R}^{} - \left(\overline{\ell_{\rm L}^{}} {\bf y}_\nu^{} \widetilde{H} N_{\rm R}^{} + \frac{1}{2} \overline{N_{\rm R}^{\rm c}} {\bf m}_{\rm R}^{} N_{\rm R}^{} + {\rm h.c.}\right) \;, 
\end{eqnarray}
where ${\cal L}_{\rm SM}^{}$ is the SM Lagrangian, $N_{\rm R}^{}$ denotes the right-handed neutrino field, $\ell_{\rm L}^{} = \left(\nu_{\rm L}^{}, l_{\rm L}^{}\right)^{\rm T}$ is the left-handed lepton doublet, and $\widetilde{H} \equiv {\rm i} \sigma^2 H^*$ with $H$ being the SM Higgs doublet. In Eq.~(\ref{eq:L_SS}), $N_{\rm R}^{\rm c} \equiv \mathsf{C} \overline{N_{\rm R}^{}}^{\rm T}$ has been defined with $\mathsf{C}$ being the charge-conjugation matrix, and the $3\times 3$ Majorana mass matrix ${\bf m}_{\rm R}^{}$ of right-handed neutrinos is symmetric. Besides, the $3\times 3$ Dirac neutrino Yukawa coupling matrix ${\bf y}_\nu^{}$ is in general arbitrary.\footnote{We adopt the same notations as those in Ref.~\cite{Huang:2025ubs}, where the $3\times3$ matrices in the fermion flavor spaces will always be denoted by bold Latin letters.} 

After the spontaneous gauge symmetry breaking, the Higgs field $\left<H\right> = (0,v_0^{}/\sqrt{2})^{\rm T}$ acquires its vacuum expectation value $v_0^{}\approx 246~{\rm GeV}$, and thus the Dirac neutrino mass matrix is given by ${\bf m}_{\rm D}^{} = {\bf y}_\nu^{} v_0^{} /\sqrt{2}$. Different from the SM, the overall mass term for neutrinos turns out to be
\begin{eqnarray}
		\label{eq:mass_term}
		{\cal L}_{\rm mass}^{} = -\frac{1}{2} \overline{\begin{pmatrix} \nu_{\rm L}^{} & N_{\rm R}^{\rm c} \end{pmatrix}} 
		\begin{pmatrix}
			{\bf 0} & {\bf m}_{\rm D}^{} \\ {\bf m}_{\rm D}^{\rm T} & {\bf m}_{\rm R}^{} 
		\end{pmatrix} 
		\begin{pmatrix}
			\nu_{\rm L}^{\rm c} \\ N_{\rm R}^{} 
		\end{pmatrix} + {\rm h.c.} \;,
	\end{eqnarray}
where the $6\times 6$ Majorana neutrino mass matrix can be diagonalized by a $6\times 6$ unitary matrix in the following way
\begin{eqnarray}
		\label{eq:diagonal_mass_matrix}
		\begin{pmatrix} {\bf V} & {\bf R} \\ {\bf S} & {\bf U} \end{pmatrix}^\dagger
		\begin{pmatrix} {\bf 0} & {\bf m}_{\rm D}^{} \\ {\bf m}_{\rm D}^{\rm T} & {\bf m}_{\rm R}^{} \end{pmatrix}
		\begin{pmatrix} {\bf V} & {\bf R} \\ {\bf S} & {\bf U} \end{pmatrix}^* 
		= 
		\begin{pmatrix} \widehat{\bf m} & {\bf 0} \\ {\bf 0} & \widehat{\bf M} \end{pmatrix} 
\end{eqnarray}	
with $\widehat{\bf m} \equiv {\rm diag}\left\{m_1^{}, m_2^{}, m_3^{}\right\}$ and $\widehat{\bf M} \equiv {\rm diag}\left\{M_1^{}, M_2^{}, M_3^{}\right\}$ representing the masses of three light Majorana neutrinos $\nu^{}_i$ (for $i = 1, 2, 3$) and three heavy Majorana neutrinos $N^{}_i$ (for $i = 1, 2, 3$), respectively. In Eq.~(\ref{eq:diagonal_mass_matrix}), four $3\times 3$ sub-matrices ${\bf V}$, ${\bf R}$, ${\bf S}$ and ${\bf U}$ should fulfill the unitarity conditions, such as ${\bf V}{\bf V}^\dagger + {\bf R}{\bf R}^\dagger = {\bf 1}$ and ${\bf V}{\bf S}^\dagger + {\bf R}{\bf U}^\dagger = {\bf 0}$. In the flavor basis, where the charged-lepton mass matrix ${\bf m}^{}_l = {\rm diag}\{m^{}_e, m^{}_\mu, m^{}_\tau \}$ is diagonal (with $m^{}_\alpha$ for $\alpha = e, \mu, \tau$ being charged-lepton masses), only the $3\times 6$ matrix ${\cal B}\equiv ({\bf V} ~ {\bf R})$ characterizes the lepton flavor mixing and appears in the neutrino interactions with weak gauge bosons and the Higgs boson.

In the forthcoming precision era of particle physics, to scrutinize theoretical predictions with experimental measurements, we have carried out the one-loop renormalization of the type-I seesaw model in the modified minimal-subtraction ($\overline{\rm MS}$) scheme in Ref.~\cite{Huang:2025ubs}. The $\overline{\rm MS}$ scheme~\cite{Bardeen:1978yd} serves as the most convenient way to decouple heavy particles from the ultraviolet (UV) complete theory and derive the renormalization-group equations (RGEs) for physical parameters in the low-energy effective theories~\cite{Weinberg:1980wa}. As for the type-I seesaw model, when the masses of heavy Majorana neutrinos are much higher than the electroweak scale $\Lambda^{}_{\rm EW} \approx 10^2~{\rm GeV}$, one can integrate heavy Majorana neutrinos out at the seesaw scale $\Lambda^{}_{\rm SS} \approx 10^{14}~{\rm GeV}$, construct the seesaw effective field theory (SEFT) with a series of effective operators of mass dimensions higher than four~\cite{Broncano:2002rw, Zhang:2021tsq, Zhang:2021jdf}, and run the physical parameters in the SEFT from the seesaw scale to the electroweak scale via their RGEs~\cite{Wang:2023bdw, Ibarra:2024tpt}. In this way, the low-energy observables can be calculated in the SEFT and readily confronted with experimental measurements. 

If the masses of heavy Majorana neutrinos are around or just above the electroweak scale, the precision tests of the type-I seesaw model should proceed in a similar way to those of the SM. For the latter, the on-shell renormalization scheme has been widely adopted, since all the physical parameters in the on-shell scheme are defined in a manner independent of the renormalization scale, and can be directly extracted from experimental observables. The on-shell renormalization of the SM has already been systematically studied in the literature~\cite{Aoki:1982ed, Bohm:1986rj, Hollik:1988ii, Denner:1991kt, Bohm:2001yx, Sirlin:2012mh, Denner:2019vbn, Huang:2023nqf}. As one can observe from the Lagrangian in Eq.~(\ref{eq:L_SS}), the type-I seesaw model differs from the SM only in the lepton sector, so the on-shell renormalization of the former resembles that of the latter to a large extent. However, we emphasize that the full renormalization of the type-I seesaw model in the on-shell scheme is necessary and nontrivial in the following aspects:
\begin{itemize}
  \item The presence of both light and heavy Majorana neutrinos in the interactions with weak gauge bosons and the Higgs boson changes the SM results. In particular, the field and mass renormalization of weak gauge bosons and the Higgs boson will receive extra contributions from massive Majorana neutrinos.
  \item The on-shell renormalization of the lepton flavor mixing matrix has never been performed before in the type-I seesaw model. Although it has been generally discussed in Ref.~\cite{Kniehl:1996bd, Pilaftsis:2002nc, Almasy:2009kn}, the gauge independence of the counterterms for the elements of the lepton flavor mixing matrix ${\cal B}$ has not been explicitly shown.
  \item The matching relations between the $\overline{\rm MS}$ and on-shell parameters in the SM have been known at the level of two or higher loops~\cite{Martin:2019lqd, Alam:2022cdv}. For the first time, we establish such relations in the type-I seesaw model at the one-loop level. Since the on-shell parameters can be extracted from experimental measurements, these relations are indispensable for the determination of the $\overline{\rm MS}$ parameters in the type-I seesaw model and for extrapolating their values to high-energy scales via RGEs.
\end{itemize}

In the on-shell renormalization scheme, we choose physical parameters to be the fine-structure constant $\alpha$, the masses of $W$ and $Z$ bosons $\left\{m_W^{},m_Z^{}\right\}$, the Higgs-boson mass $m_h^{}$, fermion masses $m_f^{}$, and the flavor mixing matrices $\left\{{\bf V}_{}^{\rm CKM}, {\bf V}, {\bf R}\right\}$ for quarks, light and heavy Majorana neutrinos, respectively. The on-shell renormalization conditions are imposed to fix the counterterms, while the renormalization of the Cabibbo-Kobayashi-Maskawa (CKM) matrix ${\bf V}^{\rm CKM}_{}$~\cite{Cabibbo:1963yz,Kobayashi:1973fv} is properly generalized and applied to the flavor mixing matrix ${\cal B}$ in the lepton sector. The gauge independence of the renormalized flavor mixing matrices is discussed. Finally, with the explicit forms of counterterms both in the $\overline{\rm MS}$ and on-shell schemes, we derive the relations between those two sets of parameters.
	
The remaining part is organized as follows. The strategy for the on-shell renormalization is outlined in Sec.~\ref{sec:renormalization}, in which the input parameters are chosen and the tadpole scheme is introduced. The renormalization conditions for physical masses and fields, and those for the lepton flavor mixing matrix, are given and discussed. The matching relations between the on-shell parameters and the $\overline{\rm MS}$ ones are explicitly derived in Sec.~\ref{sec:conversion}. We summarize our main results in Sec.~\ref{sec:sum}.
	
\section{Renormalization in the On-shell Scheme}
\label{sec:renormalization}
	
\subsection{The strategy}

The strategy for the on-shell renormalization is similar to that in the $\overline{\rm MS}$ scheme, as summarized in Ref.~\cite{Huang:2025ubs}. Some helpful comments are in order.
\begin{itemize}
\item {\it Input parameters}.\footnote{Notice that the physical parameters in the on-shell scheme in this work and those in the $\overline{\rm MS}$ scheme in Ref.~\cite{Huang:2025ubs} follow the same notations, but there should be no confusion as only the on-shell parameters are relevant in this section. Later in Sec.~\ref{sec:conversion}, we shall distinguish between these two sets of physical parameters.} In the on-shell scheme, the input parameters are set to be the electromagnetic fine-structure constant $\alpha$, the physical masses $\left\{m_W^{}, m_Z^{}, m_h^{}, m_f^{}\right\}$ of weak gauge bosons, the Higgs boson and massive fermions, the CKM matrix ${\bf V}^{\rm CKM}_{}$ for the quark flavor mixing, and the flavor mixing matrix ${\cal B} \equiv ({\bf V} ~ {\bf R})$ for leptons. 
		
	Once the input parameters are chosen, one can rewrite the bare quantities in terms of the renormalized ones and the corresponding counterterms
		\begin{eqnarray}
			e_0^{} &=& \left(1 + \delta Z_e^{}\right) e\;, \nonumber\\
			m_{W,0}^2 &=& m_W^2 + \delta m_W^2 \;, \nonumber\\
			m_{Z,0}^2 &=& m_Z^2 + \delta m_Z^2 \;, \nonumber\\
			m_{h,0}^2 &=& m_h^2 + \delta m_h^2 \;, \nonumber\\
			m_{f,0}^{} &=& m_f^{} + \delta m_f^{} \;, \nonumber \\
			{\bf V}^{\rm CKM}_{0} &=& {\bf V}^{\rm CKM} + \delta {\bf V}^{\rm CKM}_{} \;, \nonumber \\
			{\cal B}_0^{} &=& {\cal B} + \delta {\cal B} \;,
		\end{eqnarray}
		where the bare parameters are labeled by the subscript  ``0". In addition, the weak mixing angle $\theta^{}_{\rm w}$ will be defined via $c\equiv \cos\theta_{\rm w}^{} = m_W^{}/m_Z^{}$ and $s\equiv\sin\theta_{\rm w}^{}$, while the weak gauge coupling is expressed as $g^2 = 4\pi\alpha/s^2$ with $\alpha \equiv e^2/(4\pi)$. Furthermore, we also need the renormalization counterterms for physical fields to define their mass eigenstates:
		\begin{eqnarray}
				W_{0\mu}^{\pm} &=& \left(1 + \frac{1}{2} \delta Z_W^{}\right) W_\mu^\pm \;, \nonumber \\
				\begin{pmatrix} Z_{0\mu}^{} \\ A_{0\mu}^{}\end{pmatrix} &=& \begin{pmatrix} 
					1 + \displaystyle \frac{1}{2} \delta Z_{ZZ}^{} & \displaystyle \frac{1}{2} \delta Z_{ZA}^{} \\ \displaystyle \frac{1}{2} \delta Z_{AZ}^{} & 1 + \displaystyle \frac{1}{2} \delta Z_{AA}^{}
				\end{pmatrix}
				\begin{pmatrix} Z_{\mu}^{} \\ A_{\mu}^{} \end{pmatrix}\;, \nonumber \\
				h_0^{} &=& \left(1 + \frac{1}{2} \delta Z_h^{}\right) h \;, \nonumber \\
				f_{i,0}^{\rm L} &=& \left(\delta_{ij}^{} + \frac{1}{2} \delta Z_{ij}^{f,{\rm L}}\right) f_j^{\rm L} \;, \nonumber \\
				f_{i,0}^{\rm R} &=& \left(\delta_{ij}^{} + \frac{1}{2} \delta Z_{ij}^{f,{\rm R}}\right) f_j^{\rm R} \;.
		\end{eqnarray}
To render all the renormalized Green's functions to be UV-finite, one should renormalize the unphysical fields, including the ghost and Goldstone fields, in a proper way. However, the renormalization constants for unphysical fields will not affect the physical parameters and the $S$-matrix elements. For simplicity, we shall not discuss the renormalization of the unphysical sector in this work and just refer to Refs.~\cite{Aoki:1982ed,Bohm:1986rj,Denner:2019vbn} for more details. 
		
		\item {\it Tadpole corrections}. We adopt the Fleischer-Jegerlehner tadpole scheme~\cite{Fleischer:1980ub}, in which the effective Higgs potential is expanded with respect to its tree-level minimum $v^{}_0 \approx 246~{\rm GeV}$, and thus the tadpole contributions to the two-point self-energies should be included. In such a scheme, it is guaranteed that both the renormalized physical masses and the corresponding mass counterterms are independent of the gauge parameters.
		
		\item {\it Regularization scheme}. The dimensional regularization~\cite{Bollini:1972ui, tHooft:1972tcz} will be utilized to calculate one-loop Feynman diagrams and separate out the UV-divergent terms, where the space-time dimension is set
		to $d = 4 - 2\epsilon$ and the divergent term is denoted as $\Delta \equiv 1/\epsilon - \gamma_{\rm E}^{} + \ln(4\pi)$ with the Euler-Mascheroni constant $\gamma_{\rm E}^{} \approx 0.577$.
		
	\end{itemize}
	
	\subsection{Renormalization conditions}
	
The on-shell renormalization conditions fix the renormalized masses as the real parts of the poles of the corresponding propagators, which are equivalent to the zeros of the one-particle irreducible (1PI) two-point functions up to the one-loop level. Meanwhile, although it is unnecessary to perform the wave-function renormalization in order to calculate the $S$-matrix elements, it would be convenient to determine the wave-function counterterms such that there is no mixing between physical fields. In such circumstances, we may choose the renormalized propagators of on-shell particles to have residues of one. 
	
With the above on-shell conditions, the mass and wave-function counterterms can be determined with the help of the 1PI two-point functions. First, the one-loop self-energies for the gauge fields can be expressed as
	\begin{eqnarray}
		\Sigma_{\mu \nu}^{V} (p^2) =   \left(g_{\mu\nu}^{} - \frac{p_\mu^{} p_\nu^{}}{p^2}\right) \Sigma_{\rm T}^V (p^2) + \frac{p_\mu^{} p_\nu^{}}{p^2} \Sigma_{\rm L}^V (p^2) \;,
	\end{eqnarray}
	for $V=W,Z,A,AZ$, while that for the Higgs boson can be simply denoted as ${\rm i} \Sigma_{}^h (p^2)$. Then the mass and wave-function counterterms for gauge bosons and the Higgs boson are given by
	\begin{eqnarray}
		&& \delta m_W^2 = -{\rm Re}\;\Sigma_{\rm T}^W \left(m_W^2\right)\;, \quad \delta Z_W^{} = \left.{\rm Re}\;\frac{\partial \Sigma_{\rm T}^W \left(p^2\right)}{\partial p^2}\right|_{p^2_{} = m_W^2}\;, \nonumber \\
		&& \delta m_Z^2 = -{\rm Re}\;\Sigma_{\rm T}^Z \left(m_Z^2\right)\; , \quad \delta Z_{ZZ}^{} = \left.{\rm Re}\;\frac{\partial \Sigma_{\rm T}^Z \left(p^2\right)}{\partial p^2}\right|_{p^2_{} = m_Z^2}\;, \nonumber \\
		&& \delta Z_{AZ}^{} = 2 {\rm Re}\;\frac{\Sigma_{\rm T}^{AZ}\left(m_{Z}^{2}\right)}{m_{Z}^{2}}\;, \quad \delta Z_{Z A}^{} = - 2 \frac{\Sigma_{\rm T}^{A Z}(0)}{m_{Z}^{2}}\;, \quad \delta Z_{AA}^{} = \left.\frac{\partial \Sigma_{\rm T}^{A}\left(p^2\right)}{\partial p^2}\right|_{p^2 = 0} \;, \nonumber \\
		&& \delta m_h^2 = {\rm Re}\;\Sigma_{}^h \left(m_h^2\right)\;, \quad \delta Z_h^{} = - \left.{\rm Re}\;\frac{\partial \Sigma_{}^h \left(p^2\right)}{\partial p^2}\right|_{p^2_{} = m_h^2}\;.
	\end{eqnarray}
Compared with the results in Refs.~\cite{Denner:1991kt,Denner:2019vbn},	the sign difference in the gauge-boson counterterms arises from the fact that the gauge-boson self-energy is denoted as ${\rm i} \Sigma_{\rm T}^{}$ in this paper while it is $ - {\rm i} \Sigma_{\rm T}^{}$ in previous works. Notice that the real parts of the loop integrals in the self-energies are taken, but not those of the flavor mixing matrix elements. The self-energies of the photon $\Sigma_{\rm T}^{A}\left(p^2\right)$ and the $A$-$Z$ mixing $\Sigma_{\rm T}^{AZ}\left(p^{2}\right)$ are the same as those in the SM, since the electrically neutral neutrinos do not contribute to the one-loop corrections. On the other hand, the leptonic contributions due to the presence of heavy Majorana neutrinos should affect the self-energies of the gauge bosons $\Sigma_{\rm T}^{W,Z} (p^2)$ and the Higgs boson $\Sigma_{}^h \left(p^2\right)$, while the other parts remain the same as those in the SM~\cite{Marciano:1980pb,Denner:1991kt,Degrassi:1992ff,Huang:2023nqf}.
	
As for Dirac fermions, the most general form of their self-energies can be written as~\cite{Kniehl:1996bd,Pilaftsis:2002nc,Denner:2019vbn}
	\begin{eqnarray}
		\label{eq:SE_fermion}
		\Sigma_{ij}^{} (p) \equiv \slashed{p} P_{\rm L}^{} \Sigma_{ij}^{\rm L}(p^2) + \slashed{p} P_{\rm R}^{} \Sigma_{ij}^{\rm R}(p^2) + P_{\rm L}^{} \Sigma_{ij}^{\rm D}(p^2) + P_{\rm R}^{} \Sigma_{ij}^{\rm D\dagger}(p^2) \;.
	\end{eqnarray}
Then the on-shell renormalization conditions lead to the counterterms for fermion masses
	\begin{eqnarray}
		\delta m_i^{} = \frac{1}{2} m_i^{} \left[\Sigma_{ii}^{\rm L} (m_i^2) + \Sigma_{ii}^{\rm R} (m_i^2)\right] + \frac{1}{2} \left[\Sigma_{ii}^{\rm D}(m_i^2) + \Sigma_{ii}^{\rm D \dagger}(m_i^2)\right] \;,
	\end{eqnarray}
and those for the wave functions
	\begin{eqnarray}
		\delta Z_{ij}^{\rm L} &=& \frac{2}{m_i^2 - m_j^2} \left[m_j^2 \Sigma_{ij}^{\rm L} (m_j^2) + m_i^{} m_j^{} \Sigma_{ij}^{\rm R} (m_j^2) + m_i^{} \Sigma_{ij}^{\rm D} (m_j^2) + m_j^{} \Sigma_{ij}^{\rm D\dagger} (m_j^2) \right] \;, \nonumber \\
		\delta Z_{ij}^{\rm R} &=& \frac{2}{m_i^2 - m_j^2} \left[m_j^2 \Sigma_{ij}^{\rm R} (m_j^2) + m_i^{} m_j^{} \Sigma_{ij}^{\rm L} (m_j^2) + m_j^{} \Sigma_{ij}^{\rm D} (m_j^2) + m_i^{} \Sigma_{ij}^{\rm D\dagger} (m_j^2) \right] \;,
	\end{eqnarray}
	for $i\neq j$. In the case of $i = j$, we obtain the real parts of wave-function counterterms as
	\begin{eqnarray}
		{\rm Re}\,\delta Z_{ii}^{\rm L}  &=& - \Sigma_{ii}^{\rm L} (m_i^2) - m_i^2 \left[\Sigma_{ii}^{\rm L\prime} (m_i^2) + \Sigma_{ii}^{\rm R\prime} (m_i^2)\right] - m_i^{} \left[\Sigma_{ii}^{\rm D \prime}(m_i^2) + \Sigma_{ii}^{\rm D \dagger \prime}(m_i^2)\right] \;, \nonumber \\
		{\rm Re}\,\delta Z_{ii}^{\rm R}  &=& - \Sigma_{ii}^{\rm R} (m_i^2) - m_i^2 \left[\Sigma_{ii}^{\rm L\prime} (m_i^2) + \Sigma_{ii}^{\rm R\prime} (m_i^2)\right] - m_i^{} \left[\Sigma_{ii}^{\rm D \prime}(m_i^2) + \Sigma_{ii}^{\rm D \dagger \prime}(m_i^2)\right] \;,
	\end{eqnarray}
whereas the imaginary parts are not uniquely fixed but they satisfy the conditions
	\begin{eqnarray}
		m_i^{} {\rm Im} \left(\delta Z_{ii}^{\rm L} - \delta Z_{ii}^{\rm R}\right) = 2 {\rm Im}\,\Sigma_{ii}^{\rm D}(m_i^2) \;.
	\end{eqnarray}
	In the type-I seesaw model, one can always choose $\delta Z_{ii}^{\rm L,R}$ to be real by rephasing the fermion fields since $\Sigma_{ii}^{\rm D} (m_i^2)$ is real for quarks and charged leptons. However, the situation is different for Majorana neutrinos, for which we have $\delta Z_{}^{f, {\rm L}} = \delta Z_{}^{f, {\rm R*}}$ due to the Majorana condition. In this case, the corresponding mass counterterms are
	\begin{eqnarray}
		\delta m_i^{} = m_i^{} \Sigma_{ii}^{\rm L} (m_i^2) + \frac{1}{2} \left[\Sigma_{ii}^{\rm M}(m_i^2) + \Sigma_{ii}^{\rm M *}(m_i^2)\right] \;.
	\end{eqnarray}
	The off-diagonal wave-function counterterms for $i \neq j$ are expressed as
	\begin{eqnarray}
		\delta Z_{ij}^{\rm L} &=& \delta Z_{ij}^{\rm R*} = \frac{2}{m_i^2 - m_j^2} \left[m_j^2 \Sigma_{ij}^{\rm L} (m_j^2) + m_i^{} m_j^{} \Sigma_{ij}^{\rm R} (m_j^2) + m_i^{} \Sigma_{ij}^{\rm M} (m_j^2) + m_j^{} \Sigma_{ij}^{\rm M *} (m_j^2) \right] \;,
	\end{eqnarray}
	while for $i=j$ we further have
	\begin{eqnarray}
		\delta Z_{ii}^{\rm L} = - \Sigma_{ii}^{\rm L} (m_i^2) - 2 m_i^2 \Sigma_{ii}^{\rm L\prime} (m_i^2) - m_i^{} \left[\Sigma_{ii}^{\rm M \prime}(m_i^2) + \Sigma_{ii}^{\rm M *\prime}(m_i^2)\right] + \frac{\Sigma_{ii}^{\rm M}(m_i^2) - \Sigma_{ii}^{\rm M *}(m_i^2)}{2 m_i^{}} \;. 
	\end{eqnarray}
	There is no degree of freedom for the phase redefinition in the Majorana case, so the diagonal wave-function counterterms are also fixed. With the specific expressions of the self-energy corrections for the quarks, charged leptons and Majorana neutrinos (see, e.g., Refs.~\cite{Denner:1991kt,Pilaftsis:2002nc,Huang:2025ubs}), one may directly obtain the mass and wave-function renormalization constants.
	 
\subsection{Renormalization of the Mixing Matrices}
	
With the fermion wave-function counterterms in the previous subsection, we are ready to derive the counterterms for the flavor mixing matrices. It has been shown that only the anti-Hermitian parts of the left-handed wave-function counterterms contribute to the renormalization of the mixing matrices. For example, for the CKM matrix, we have the counterterm~\cite{Denner:1990yz}
	\begin{eqnarray}
		\label{eq:delta_VCKM}
		\widetilde{\delta} {\bf V}^{\rm CKM}_{} = - \frac{1}{4} \left[{\bf V}^{\rm CKM}_{} \left(\delta Z_{}^{d,{\rm L}} - \delta Z_{}^{d,{\rm L}\dagger}\right) + \left(\delta Z_{}^{u,{\rm L}\dagger} - \delta Z_{}^{u,{\rm L}}\right) {\bf V}^{\rm CKM}_{}\right] \;.
	\end{eqnarray}
In the lepton sector, the $3 \times 3$ matrix ${\bf V}$, which is the  Pontecorvo-Maki-Nakagawa-Sakata (PMNS) matrix~\cite{Pontecorvo:1957cp,Maki:1962mu,Pontecorvo:1967fh}, and ${\bf R}$ are physical. Therefore, we need to consider the renormalization of the $3\times 6$ matrix ${\cal B}$. Similarly, we arrive at~\cite{Kniehl:1996bd,Pilaftsis:2002nc}
	\begin{eqnarray}
		\label{eq:delta_B}
		\widetilde{\delta} {\cal B} = - \frac{1}{4} \left[{\cal B} \left(\delta Z_{}^{\chi,{\rm L}} - \delta Z_{}^{\chi,{\rm L}\dagger}\right) + \left(\delta Z_{}^{l,{\rm L}\dagger} - \delta Z_{}^{l,{\rm L}}\right) {\cal B}\right] \;,
	\end{eqnarray}
where the superscript ``$\chi$" refers to massive Majorana neutrinos $\nu^{}_i$ and $N^{}_i$ (for $i  =1, 2, 3$). However, the counterterms calculated from Eqs.~(\ref{eq:delta_VCKM}) and (\ref{eq:delta_B}) are actually gauge-dependent. This is the reason why those counterterms are denoted by  $\widetilde{\delta}{\bf V}^{\rm CKM}_{}$ and $\widetilde{\delta} {\cal B}$ instead of ${\delta}{\bf V}^{\rm CKM}_{}$ and ${\delta} {\cal B}$. 

For the CKM matrix, several specific modified schemes have been proposed in the literature to derive the gauge-independent counterterms, and the phenomenological impacts have also been discussed~\cite{Gambino:1998ec,Kniehl:2000rb,Yamada:2001px,Liao:2003jy,Kniehl:2006bs,Kniehl:2006rc,Almasy:2008ep,Kniehl:2009kk,Almasy:2011vy}. Nevertheless, all those schemes cannot be directly applied to the lepton sector, where massive Majorana neutrinos are present. Although the diagonal wave-function counterterms do not contribute to the counterterm of the mixing matrix in the case of Dirac fermions, they are relevant in the case of Majorana neutrinos and have not yet been fully taken into consideration. Therefore, in the following part of this subsection, we propose a practical method to determine the gauge-independent counterterms of the leptonic flavor mixing matrix:
	\begin{itemize}
		\item First, starting with the fermion wave-function counterterms $\{\delta Z_{}^{\chi,{\rm L}}, \delta Z_{}^{l,{\rm L}}\}$ fixed through the on-shell conditions, one can obtain the explicit expression of Eq.~(\ref{eq:delta_B}) but now rewrite it as $\widetilde{\delta}{\cal B} \equiv \widetilde{\delta}{\cal B}_{}^l +  \widetilde{\delta}{\cal B}_{}^\chi$, where $\widetilde{\delta}{\cal B}_{}^l$ and $\widetilde{\delta}{\cal B}_{}^\chi$ refer to the anti-Hermitian parts of charged leptons and Majorana neutrinos in Eq.~(\ref{eq:delta_B}), respectively. 
		
		\item Then, for charged leptons, based on the expression of $\widetilde{\delta}{\cal B}_{}^l$, one can further reduce all the relevant Passarino-Veltman (PV) functions~\cite{Passarino:1978jh} into the scalar one- and two-point functions. In this way, we can divide the whole expression into two parts according to whether the gauge parameters are explicitly involved in the scalar loop functions, i.e., $\widetilde{\delta}{\cal B}_{}^{l} \equiv \widetilde{\delta}{\cal B}_{}^{l,{\rm GI}} + \widetilde{\delta}{\cal B}_{}^{l,\xi}$, where $\widetilde{\delta}{\cal B}_{}^{l,{\rm GI}}$ is gauge-independent (GI) and $\widetilde{\delta}{\cal B}_{}^{l,\xi}$ represents the part with the gauge parameter $\xi^{}_W$ or $\xi^{}_Z$ involved. To be more explicit, after reducing all loop functions in self-energies into scalar one- and two-point functions, we obtain the gauge-independent terms
		\begin{eqnarray}
			\widetilde{\delta}{\cal B}_{\alpha i}^{l,{\rm GI}} &=& - \frac{\alpha }{32 \pi s^2 m_W^2} \sum_{\beta\neq \alpha} \sum_k \frac{{\cal B}_{\alpha k}^{} {\cal B}_{\beta k}^* {\cal B}_{\beta i}^{}}{m_{\alpha}^2-m_{\beta}^2} \left\{ \vphantom{\frac{1}{1}} \left[2 \widehat{m}_k^2+m_{\alpha}^2+m_{\beta }^2+2 (d-2) m_W^2 \right] A_0^{}\left(\widehat{m}_k^{}\right) \right.  \nonumber \\
			&& - \left\{\left[ \left(\widehat{m}_k^2 - m_{\alpha}^2\right)^2 + (d-3) \left(\widehat{m}_k^2 + m_\alpha^2 \right) m_W^2 -(d-2) m_W^4 \right] B_0^{}\left(m_{\alpha}^2;\widehat{m}_k^{},m_W^{}\right) \right. \nonumber \\
			&& \left.\left. + \left(\alpha \to \beta \right) \vphantom{\left[ \left(\widehat{m}_k^2\right)^2 \right]} \right\} \vphantom{\frac{1}{1}}\right\} \;,
		\end{eqnarray}
		with the scalar one-point function
		\begin{eqnarray}
			A_0^{}(m) = m^2 \left[\Delta + \ln\left(\frac{\mu ^2}{m^2}\right) +1\right] \;,
		\end{eqnarray}
		and the two-point function
		\begin{eqnarray}
			B_0^{} \left(p^2;m_\alpha^{},m_\beta^{}\right) = \Delta +\ln\left(\frac{\mu^2}{m_\beta^2}\right)  - \frac{m_\alpha^2-m_\beta^2+p^2}{2 p^2} \ln\left(\frac{m_\alpha^2}{m_\beta^2}\right)  + \Lambda\left(p^2;m_\alpha^{},m_\beta^{}\right) + 2 \;, \qquad
		\end{eqnarray}
		where we define the function $\Lambda$ as
		\begin{eqnarray}
			\Lambda\left(p^2;m_\alpha^{},m_\beta^{}\right) = \frac{\sqrt{\lambda\left(p^2,m_\alpha^2,m_\beta^2\right)}}{p^2} \ln \left[\frac{\sqrt{\lambda\left(p^2,m_\alpha^2,m_\beta^2\right)}+m_\alpha^2+m_\beta^2-p^2}{2 m_\alpha^{} m_\beta^{}}\right] \;,
		\end{eqnarray}
with the K\"{a}ll\'{e}n function $\lambda(x,y,z) \equiv x^2 + y^2 + z^2 - 2xy - 2xz - 2yz$. Note that the 't Hooft mass scale $\mu$ has been introduced as usual. Meanwhile, all terms with gauge parameters in the loop functions are 
		\begin{eqnarray}
			\widetilde{\delta}{\cal B}_{\alpha i}^{l,\xi} &=& \frac{\alpha}{32 \pi s^2 m_W^2} \sum_{\beta\neq \alpha} \sum_k {\cal B}_{\alpha k}^{} {\cal B}_{\beta k}^* {\cal B}_{\beta i}^{} \nonumber \\
			&& \times \left[ \left(\widehat{m}_k^2-m_{\alpha}^2+\xi_W^{} m_W^2 \right) B_0^{}\left(m_{\alpha}^2;\widehat{m}_k^{},\sqrt{\xi_W^{}} m_W^{} \right) - \left(\alpha \to \beta \right) \right] \;.
		\end{eqnarray}
One can immediately verify that it is UV-finite with the help of the unitarity condition
		\begin{eqnarray}
			\sum_k {\cal B}_{\alpha k}^{} {\cal B}_{\beta k}^* = \delta_{\alpha\beta} \;.
		\end{eqnarray}
On the other hand, for the charged leptons, $\Sigma_{\alpha\beta}^{\rm L,R}$ in their self-energies are the elements of Hermitian matrices and $\Sigma_{\alpha\alpha}^{\rm D}$ are real, so there are no contributions from diagonal parts of the wave functions. Only the gauge-independent part $\widetilde{\delta}{\cal B}_{}^{l,{\rm GI}}$ will be incorporated into the counterterm of the flavor mixing matrix. The UV-finite but gauge-dependent part $\widetilde{\delta}{\cal B}_{}^{l,\xi}$ enters into the one-loop amplitude for a specific physical process. However, it is generally expected that the dependence on the gauge parameters in the observables should be canceled out after including such terms. 
		
		\item For Majorana neutrinos, we also make the decomposition $\widetilde{\delta}{\cal B}_{}^{\chi} \equiv \widetilde{\delta}{\cal B}_{}^{\chi,{\rm GI}} + \widetilde{\delta}{\cal B}_{}^{\chi,\xi}$ in the same way, with the gauge-independent part
		\begin{eqnarray}
			\widetilde{\delta}{\cal B}_{\alpha i}^{\chi,{\rm GI}} &=& -\frac{\alpha {\cal B}_{\alpha i}^{} }{64 \pi s^2 \widehat{m}_i^{} m_W^2} \sum_k \widehat{m}_k^{} \left({\cal C}_{ki}^2 - {\cal C}_{ik}^2 \right) \nonumber \\
			&& \times \left[ \left(\widehat{m}_k^2-\widehat{m}_i^2\right) B_0^{}\left(\widehat{m}_i^2;m_h^{},\widehat{m}_k^{}\right) +(d-1) m_Z^2 B_0^{}\left(\widehat{m}_i^2;\widehat{m}_k^{},m_Z^{}\right) \vphantom{\left(\sqrt{\xi_Z^{}}\right)} \right] \nonumber \\
			&& + \frac{\alpha}{32 \pi s^2 m_W^2} \sum_{j \neq i} \frac{{\cal B}_{\alpha j}^{}}{\widehat{m}_i^{} \widehat{m}_j^{} \left(\widehat{m}_j^2-\widehat{m}_i^2\right)} \nonumber \\
			&& \times \left\{ \sum_\rho \left\{ {\cal B}_{\rho j}^{} {\cal B}_{\rho i}^* \left[2 \widehat{m}_i^2\widehat{m}_j^2 + (d-2) m_W^2 \left(\widehat{m}_i^2 + \widehat{m}_j^2 \right) + \left(\widehat{m}_i^2 + \widehat{m}_j^2\right) m_{\rho}^2\right] \right. \right. \nonumber \\
			&& \left. + {\cal B}_{\rho j}^* {\cal B}_{\rho i}^{} \widehat{m}_i^{} \widehat{m}_j^{} \left[\widehat{m}_i^2+\widehat{m}_j^2+ 2(d-2) m_W^2+2 m_{\rho}^2\right] \right\} A_0^{}\left(m_{\rho}^{}\right) \nonumber \\
			&& - \left\{ \sum_\rho \left[{\cal B}_{\rho j}^{} {\cal B}_{\rho i}^* m_{\rho}^2 \left(\widehat{m}_i^2+\widehat{m}_j^2\right) + 2 {\cal B}_{\rho j}^* {\cal B}_{\rho i}^{} m_{\rho}^2 \widehat{m}_i^{} \widehat{m}_j^{}\right]   \right. \nonumber \\
			&& - {\cal C}_{ji}^{} \widehat{m}_i^{} \widehat{m}_j^{} \left[\widehat{m}_i^2+\widehat{m}_j^2-2 (d-2) m_W^2\right] - {\cal C}_{ji}^* \left[2 \widehat{m}_i^2 \widehat{m}_j^2 - (d-2)  \left(\widehat{m}_i^2+\widehat{m}_j^2\right) m_W^2\right]  \nonumber \\
			&& \left. - 4 (d-1) \widehat{m}_i^{} \widehat{m}_j^{} \frac{m_W^2}{m_h^2} \left[{\cal C}_{ji}^{} \left(\widehat{m}_i^2+\widehat{m}_j^2\right)+2 \widehat{m}_i^{} \widehat{m}_j^{} {\cal C}_{ji}^* \right] \right\} A_0^{}\left(m_W^{}\right)  \nonumber \\
			&& +  \left\{ \vphantom{\sum_k} 5 {\cal C}_{ji}^* \widehat{m}_i^2  \widehat{m}_j^2 + \frac{5}{2} {\cal C}_{ji}^{} \widehat{m}_i^{} \widehat{m}_j^{} \left(\widehat{m}_i^2+\widehat{m}_j^2\right)  \right. \nonumber \\
			&& \left. - \frac{1}{2} \sum_k  \left[{\cal C}_{jk}^* {\cal C}_{ki}^* \widehat{m}_k^2 \left(\widehat{m}_i^2 + \widehat{m}_j^2\right) + 2 {\cal C}_{jk} {\cal C}_{ki} \widehat{m}_i^{} \widehat{m}_j^{} \widehat{m}_k^2\right]  \right\} A_0^{}(m_h^{}) \nonumber \\
			&& -\frac{\widehat{m}_j^{}}{2} \sum_k \left\{ {\cal C}_{jk}^{} {\cal C}_{ki}^{} \widehat{m}_i^{} \left[\widehat{m}_k^4 + \left(3 \widehat{m}_i^2 + 3 \widehat{m}_j^2 - m_h^2\right) \widehat{m}_k^2  + \left(\widehat{m}_i^2 - m_h^2\right) \widehat{m}_j^2 \right] \right.  \nonumber \\
			&&  + {\cal C}_{jk}^{} {\cal C}_{ki}^* \widehat{m}_k \left[\widehat{m}_i^4+\left(3 \widehat{m}_j^2 + 3 \widehat{m}_k^2 - m_h^2\right) \widehat{m}_i^2+ \left(\widehat{m}_k^2-m_h^2\right)\widehat{m}_j^2\right] \nonumber \\
			&& + {\cal C}_{jk}^* {\cal C}_{ki}^* \widehat{m}_j^{} \left[\widehat{m}_i^4-\left(m_h^2-6 \widehat{m}_k^2\right) \widehat{m}_i^2+\widehat{m}_k^4-m_h^2 \widehat{m}_k^2\right] \nonumber \\
			&& \left. +  2 {\cal C}_{jk}^* {\cal C}_{ki}^{} \widehat{m}_i^{} \widehat{m}_j^{} \widehat{m}_k^{} \left(2 \widehat{m}_i^2+2 \widehat{m}_k^2-m_h^2\right) \right\} B_0^{}\left(\widehat{m}_i^2;m_h^{},\widehat{m}_k^{}\right)  \nonumber \\
			&& -\frac{\widehat{m}_i^{}}{2} \sum_k \left\{ {\cal C}_{jk}^{} {\cal C}_{ki}^{} \widehat{m}_j^{} \left[\widehat{m}_k^4 + \left(3 \widehat{m}_i^2 + 3 \widehat{m}_j^2 - m_h^2\right) \widehat{m}_k^2  + \left(\widehat{m}_j^2 - m_h^2\right) \widehat{m}_i^2 \right] \right. \nonumber \\
			&&  +  {\cal C}_{jk}^* {\cal C}_{ki}^{} \widehat{m}_k^{} \left[\widehat{m}_j^4+\left(3 \widehat{m}_i^2 + 3 \widehat{m}_k^2 - m_h^2\right) \widehat{m}_j^2+ \left(\widehat{m}_k^2-m_h^2\right)\widehat{m}_i^2\right] \nonumber \\
			&& + {\cal C}_{jk}^* {\cal C}_{ki}^* \widehat{m}_i \left[\widehat{m}_j^4-\left(m_h^2-6 \widehat{m}_k^2\right) \widehat{m}_j^2+\widehat{m}_k^4-m_h^2 \widehat{m}_k^2\right] \nonumber \\
			&& \left. + 2 {\cal C}_{jk}^{} {\cal C}_{ki}^* \widehat{m}_i^{} \widehat{m}_j^{}  \widehat{m}_k^{} \left(2 \widehat{m}_j^2+2 \widehat{m}_k^2-m_h^2\right) \right\} B_0^{}\left(\widehat{m}_j^2;m_h^{},\widehat{m}_k^{}\right) \nonumber \\
			&& +\frac{1}{2} \sum_k \left\{ 2 {\cal C}_{jk}^{} {\cal C}_{ki}^{} \widehat{m}_i^{} \widehat{m}_j^{} \left[\widehat{m}_i^2+\widehat{m}_j^2+2 \widehat{m}_k^2+(d-2) m_Z^2\right] \right. \nonumber \\
			&& + 2 {\cal C}_{jk}^{} {\cal C}_{ki}^* \widehat{m}_j^{} \widehat{m}_k^{} \left(3 \widehat{m}_i^2+\widehat{m}_j^2\right) +  2 {\cal C}_{jk}^* {\cal C}_{ki}^{} \widehat{m}_i^{} \widehat{m}_k^{} \left(3 \widehat{m}_j^2+ \widehat{m}_i^2\right) \nonumber \\
			&& \left. + {\cal C}_{jk}^* {\cal C}_{ki}^* \left[4 \widehat{m}_i^2 \widehat{m}_j^2+2 \left(\widehat{m}_i^2+\widehat{m}_j^2\right) \widehat{m}_k^2+(d-2) \left(\widehat{m}_i^2+\widehat{m}_j^2\right) m_Z^2\right] \right\} A_0^{}\left(\widehat{m}_k^{}\right)  \nonumber \\
			&& -\frac{\widehat{m}_j^{}}{2} \sum_k \left\{\vphantom{\frac{1}{1}} 2 (d-1) \widehat{m}_i^{} \widehat{m}_k^{} m_Z^2 \left[\widehat{m}_i^{} {\cal C}_{jk}^{} {\cal C}_{ki}^* + \widehat{m}_j^{} {\cal C}_{jk}^* {\cal C}_{ki}^{}\right]\right. \nonumber \\
			&& + \left(\widehat{m}_i^{} {\cal C}_{jk}^{} {\cal C}_{ki}^{} + \widehat{m}_j^{} {\cal C}_{jk}^*{\cal C}_{ki}^* \right) \times \left\{\widehat{m}_i^4 + \left[(d-3) m_Z^2-2 \widehat{m}_k^2\right] \widehat{m}_i^2 \right.\nonumber \\
			&& \left. \left.  +\left(\widehat{m}_k^2-m_Z^2\right) \left[\widehat{m}_k^2+(d-2) m_Z^2\right] \right\} \vphantom{\frac{1}{1}} \right\} B_0^{}\left(\widehat{m}_i^2;\widehat{m}_k^{},m_Z^{}\right)  \nonumber \\
			&& -\frac{\widehat{m}_i^{}}{2} \sum_k \left\{\vphantom{\frac{1}{2}} 2 (d-1) \widehat{m}_j^{} \widehat{m}_k^{} m_Z^2 \left[\widehat{m}_i^{} {\cal C}_{jk}^{} {\cal C}_{ki}^* + \widehat{m}_j^{} {\cal C}_{jk}^*{\cal C}_{ki}^{}  \right]  \right. \nonumber \\
			&& + \left(\widehat{m}_j^{} {\cal C}_{jk}^{} {\cal C}_{ki}^{}  + \widehat{m}_i^{} {\cal C}_{jk}^* {\cal C}_{ki}^*  \right) \times  \left\{\widehat{m}_j^4+\left[(d-3) m_Z^2-2 \widehat{m}_k^2\right] \widehat{m}_j^2 \right. \nonumber \\
			&& \left. \left.  + \left(\widehat{m}_k^2-m_Z^2\right) \left[\widehat{m}_k^2 + (d-2) m_Z^2 \right] \right\} \vphantom{\frac{1}{2}}\right\} B_0^{}\left(\widehat{m}_j^2;\widehat{m}_k^{},m_Z^{}\right)  \nonumber \\
			&& +\frac{1}{2} \sum_k \left\{ 4 (d-1) \widehat{m}_i^{} \widehat{m}_j^{} \frac{m_Z^2}{m_h^2} \left[{\cal C}_{ji}^{} \left(\widehat{m}_i^2+\widehat{m}_j^2\right)+2 \widehat{m}_i^{} \widehat{m}_j^{} {\cal C}_{ji}^* \right]  \right. \nonumber \\
			&& -2 {\cal C}_{jk}^{} {\cal C}_{ki}^{} \widehat{m}_i^{} \widehat{m}_j^{} \widehat{m}_k^2 - {\cal C}_{jk}^* {\cal C}_{ki}^* \left(\widehat{m}_i^2+\widehat{m}_j^2\right) \widehat{m}_k^2+{\cal C}_{ji}^{} \widehat{m}_i^{} \widehat{m}_j^{} \left[\widehat{m}_i^2+\widehat{m}_j^2-2 (d-2) m_Z^2\right]  \nonumber \\
			&& \left. + {\cal C}_{ji}^* \left[2 \widehat{m}_i^2 \widehat{m}_j^2-(d-2) \left(\widehat{m}_i^2+\widehat{m}_j^2\right) m_Z^2\right]   \vphantom{\frac{1}{2}} \right\} A_0^{}\left(m_Z^{}\right)  \nonumber \\
			&& - 8 \widehat{m}_i^{} \widehat{m}_j^{} \left[{\cal C}_{ji} \left(\widehat{m}_i^2 + \widehat{m}_j^2 \right) + 2 \widehat{m}_i^{} \widehat{m}_j^{} {\cal C}_{ji}^* \right] \sum_{f=q,l} N_{\rm c}^f \frac{m_f^2}{m_h^2} A_0^{}\left(m_f^{}\right)   \nonumber \\
			&& - 8 \widehat{m}_i^{} \widehat{m}_j^{} \left[{\cal C}_{ji} \left(\widehat{m}_i^2 + \widehat{m}_j^2 \right) + 2 \widehat{m}_i^{} \widehat{m}_j^{} {\cal C}_{ji}^* \right] \sum_{k} {\cal C}_{kk}^{} \frac{\widehat{m}_k^2}{m_h^2} A_0^{}\left(\widehat{m}_k^{}\right)  \nonumber \\
			&& -\widehat{m}_j^{} \sum_\rho \left(\widehat{m}_j^{} {\cal B}_{\rho j}^{} {\cal B}_{\rho i}^* + \widehat{m}_i^{} {\cal B}_{\rho j}^* {\cal B}_{\rho i}^{} \right) \times \left[\widehat{m}_i^4 +m_{\rho}^4 \right. \nonumber \\
			&& \left.  + (d-3) m_W^2 \left(\widehat{m}_i^2 + m_{\rho}^2 \right) - 2 m_{\rho}^2 \widehat{m}_i^2 - (d-2) m_W^4 \right] B_0^{}\left(\widehat{m}_i^2;m_W^{},m_{\rho }^{}\right) \nonumber \\
			&& - \widehat{m}_i^{} \sum_\rho \left(\widehat{m}_i^{} {\cal B}_{\rho j}^{} {\cal B}_{\rho i}^* + \widehat{m}_j^{} {\cal B}_{\rho j}^* {\cal B}_{\rho i}^{}  \right) \times \left[\widehat{m}_j^4 + m_{\rho}^4 \right. \nonumber \\
			&& \left.  \left. + (d-3) m_W^2 \left(\widehat{m}_j^2 + m_\rho^2\right) - 2 m_{\rho}^2 \widehat{m}_j^2 - (d-2) m_W^4 \right] B_0^{}\left(\widehat{m}_j^2;m_W^{},m_{\rho}^{}\right) \vphantom{\sum_\rho}\right\} \;,
		\end{eqnarray}
where $\widehat{m}^{}_i$ (for $i = 1, 2, \cdots, 6$) run over six Majorana neutrino masses with the identification $\widehat{m}^{}_i = m^{}_i$ and $\widehat{m}^{}_{i+3} = M^{}_i$ (for $i = 1, 2, 3$) and ${\cal C} \equiv {\cal B}^\dagger {\cal B}$ is a $6\times 6$ Hermitian matrix, and $N_{\rm c}^f$ is the color factor with $N_{\rm c}^q=3$ for quarks and $N_{\rm c}^l=1$ for leptons. The other part involving gauge parameters in the scalar functions is
		\begin{eqnarray}
			\widetilde{\delta}{\cal B}_{\alpha i}^{\chi,\xi} &=& -\frac{\alpha {\cal B}_{\alpha i}^{} }{64 \pi s^2 \widehat{m}_i^{} m_W^2} \sum_k \widehat{m}_k^{} \left(\widehat{m}_i^2-\widehat{m}_k^2+\xi_Z^{} m_Z^2 \right) \left({\cal C}_{ki}^2 - {\cal C}_{ik}^2 \right) B_0^{}\left(\widehat{m}_i^2;\widehat{m}_k^{},\sqrt{\xi_Z^{}} m_Z^{} \right)  \nonumber \\
			&& + \frac{\alpha}{32 \pi s^2 m_W^2} \sum_{j \neq i} {\cal B}_{\alpha j}^{}  \left\{\sum_{\rho} {\cal B}_{\rho j}^* {\cal B}_{\rho i}^{} \right. \nonumber \\
			&& \times \left[ \left(\widehat{m}_j^2 - \xi_W m_W^2 - m_\rho^2 \right) B_0^{}\left(\widehat{m}_j^2;m_\rho^{},\sqrt{\xi_W^{}} m_W^{} \right) - \left(j \to i \right) \right]  \nonumber \\
			&& + \frac{1}{2 \widehat{m}_j^{} } \sum_k \left[{\cal C}_{jk}^{} {\cal C}_{ki}^{} \widehat{m}_j^{} \left(\widehat{m}_j^2-\widehat{m}_k^2-\xi_Z^{} m_Z^2\right) \right. \nonumber \\
			&& \left. - {\cal C}_{jk}^* {\cal C}_{ki}^{} \widehat{m}_k^{} \left(\widehat{m}_j^2-\widehat{m}_k^2+\xi_Z^{} m_Z^2\right)\right] B_0^{}\left(\widehat{m}_j^2;\widehat{m}_k^{},\sqrt{\xi_Z^{}} m_Z^{}\right)  \nonumber \\
			&& - \frac{1}{2 \widehat{m}_i^{} } \sum_k\left[ {\cal C}_{jk}^{} {\cal C}_{ki}^{} \widehat{m}_i^{} \left(\widehat{m}_i^2-\widehat{m}_k^2-\xi_Z^{} m_Z^2 \right) \right. \nonumber \\
			&& \left. \left.  - {\cal C}_{jk}^{} {\cal C}_{ki}^* \widehat{m}_k^{} \left(\widehat{m}_i^2-\widehat{m}_k^2+\xi_Z^{} m_Z^2\right) \right] B_0^{}\left(\widehat{m}_i^2;\widehat{m}_k^{},\sqrt{\xi_Z^{}} m_Z^{}\right) \vphantom{\sum_{\rho}} \right\} \;.
		\end{eqnarray}
Notice that, different from Dirac fermions, both the diagonal and off-diagonal wave-function counterterms of Majorana neutrinos contribute. In contrast, there are also UV divergences in $\widetilde{\delta}{\cal B}_{\alpha i}^{\chi,\xi}$, namely, the terms proportional to $\Delta$ as below
		\begin{eqnarray}
			\left[\widetilde{\delta}{\cal B}_{\alpha i}^{\chi,\xi}\right]_{\rm div}^{} &=&  \frac{\alpha \Delta  }{64 \pi s^2 m_W^2} \sum_{j \neq i} {\cal B}_{\alpha j}^{} \left[3 {\cal C}_{ji}^{} \left(\widehat{m}_j^2-\widehat{m}_i^2\right)+ \sum_k \frac{\widehat{m}_k^3}{\widehat{m}_i^{} \widehat{m}_j^{}} \left({\cal C}_{jk}^* {\cal C}_{ki}^{} \widehat{m}_i^{} - {\cal C}_{jk}^{} {\cal C}_{ki}^* \widehat{m}_j^{} \right)\right]  \nonumber \\
			&& + \frac{\alpha \Delta {\cal B}_{\alpha i}^{}}{64 \pi s^2  m_W^2} \sum_k \frac{\widehat{m}_k^3}{\widehat{m}_i^{}} \left({\cal C}_{ki}^2 - {\cal C}_{ik}^2 \right) \;,
		\end{eqnarray}
where the subscript ``div" on the left-hand side refers to the UV-divergent parts in the final results. In the derivation of the above equation, we have made use of the following identity
		\begin{eqnarray}
			\sum_k \widehat{m}_k^{} {\cal C}_{ik} {\cal C}_{kj}^* = 0 \;,
		\end{eqnarray}
to demonstrate that such UV-divergent terms are independent of the gauge parameters. Therefore, for Majorana neutrinos, apart from $\widetilde{\delta}{\cal B}_{\alpha i}^{\chi,{\rm GI}}$, the UV-divergent term $[\widetilde{\delta}{\cal B}_{\alpha i}^{\chi,\xi}]_{\rm div}^{}$ should also be included in the counterterm of lepton flavor mixing matrix. As a result, the counterterm of the flavor mixing matrix remains gauge-independent and contains all the UV divergences from the anti-Hermitian combination in Eq.~(\ref{eq:delta_B}). 
	\end{itemize}
	
To summarize, we arrive at the gauge-independent counterterm of the leptonic flavor mixing matrix, namely,
	\begin{eqnarray}
		\delta {\cal B}_{}^{} = \widetilde{\delta}{\cal B}_{}^{l,{\rm GI}} + \widetilde{\delta}{\cal B}_{}^{\chi,{\rm GI}} + \left[\widetilde{\delta}{\cal B}_{}^{\chi,\xi}\right]_{\rm div}^{} \;.
	\end{eqnarray}
	The above procedure can be regarded as a generalization of the suggested renormalization of the CKM matrix in Ref.~\cite{Liao:2003jy}, where the counterterm is given by
	\begin{eqnarray}
		\delta {\bf V}_{}^{\rm CKM} = \left[\widetilde{\delta}{\bf V}_{}^{\rm CKM}\right]^{u,{\rm GI}}_{} + \left[\widetilde{\delta}{\bf V}_{}^{\rm CKM}\right]^{d,{\rm GI}}_{} \;.
	\end{eqnarray}
	However, it is worthwhile to stress that the renormalization of the mixing matrix in the lepton sector is nontrivial and very different from that in the quark sector. On the one hand, due to the Majorana nature of massive neutrinos, there are also contributions from the diagonal parts of neutrino wave-function counterterms, which come from the scalar part of diagonal self-energies $\Sigma_{ii}^{\rm M}$. On the other hand, one should include all the UV-divergent terms from $\widetilde{\delta}{\cal B}_{}^{\chi,\xi}$ into the counterterm. To the best of our knowledge, such a practical procedure and the resultant counterterm in the on-shell scheme for the lepton flavor mixing matrix have not been discussed in the previous literature.
	
Before ending this subsection, we shall emphasize that such a modified on-shell scheme for the renormalization of the lepton flavor mixing matrix is self-consistent. As a concrete example, we consider the leptonic decays of the $W$-boson, i.e., $W^- \to l_\alpha^{-} + \chi_i^{}$, with $m_W^{} \gg m_\alpha^{},\widehat{m}_i^{}$. At the leading order, the decay amplitude reads
	\begin{eqnarray}
		\mathscr{M}_0^{} = \frac{{\rm i} e}{\sqrt{2} s} \overline{u_\alpha^{}}(p_\alpha^{}) \slashed{\varepsilon}_W^{}(p_W^{}) P_{\rm L}^{} {\cal B}_{\alpha i}^{} v_i^{}(p_i^{}) \;, 
	\end{eqnarray}
	with $u_\alpha^{}(p_\alpha^{})$ and $v_i^{}(p_i^{})$ being the spinor wave function of the charged lepton and the Majorana neutrino, respectively, and $\varepsilon_W^{}(p_W^{})$ being the polarization vector of the $W$-boson. At the one-loop level, the total decay amplitude is $\mathscr{M}_1^{} = \mathscr{M}_0^{} + \mathscr{M}_{\rm vertex}^{} + \mathscr{M}_{\rm CT}^{}$, where the one-loop vertex corrections are collected in $\mathscr{M}_{\rm vertex}^{}$, and
	\begin{eqnarray}
		\label{eq:WdecayM1}
		\mathscr{M}_{\rm CT}^{} &=& \mathscr{M}_0^{} \left[\delta Z_e^{} - \frac{1}{2s^2}\frac{m_W^2}{m_Z^2}\left(\frac{\delta m_Z^2}{m_Z^2} - \frac{\delta m_W^2}{m_W^2}\right) + \frac{1}{2} \delta Z_W^{} \right] {\cal B}_{\alpha i}^{}  \nonumber \\
		&& + \mathscr{M}_0^{} \left(\frac{1}{2} \sum_\beta \delta Z_{\alpha\beta}^{l,{\rm L}\dagger} {\cal B}_{\beta i}^{} + \frac{1}{2} \sum_j {\cal B}_{\alpha j}^{} \delta Z_{ji}^{\chi,{\rm L}} \right) + \mathscr{M}_0^{} \delta {\cal B}_{\alpha i}^{} 
	\end{eqnarray}
denotes the contribution from the counterterm Lagrangian. Through the on-shell renormalization conditions, it is guaranteed that the total amplitude $\mathscr{M}_1^{}$ must be UV-finite. Furthermore, the counterterms of the coupling constant $\delta Z_e^{}$, gauge-boson masses $\left\{\delta m_W^2, \delta m_Z^2\right\}$, the flavor mixing matrix elements $\delta {\cal B}_{\alpha i}^{}$ and the renormalized amplitude $\mathscr{M}_1^{}$ are expected to be independent of gauge parameters. This indicates that the dependence on gauge parameters in $\mathscr{M}_{\rm vertex}^{}$ must cancel out with that from $\{\delta Z_W^{}, \delta Z_{\alpha\beta}^{l,{\rm L}}, \delta Z_{ji}^{\chi,{\rm L}}\}$ in $\mathscr{M}_{\rm CT}^{}$. Now, with the gauge-independent counterterm $\delta {\cal B}_{\alpha i}^{}$ determined in our practical procedure, only the UV-divergences in the anti-Hermitian parts of $\delta Z_{\alpha\beta}^{l,{\rm L}}$ and $\delta Z_{ji}^{\chi,{\rm L}}$ are eliminated, while the structure of their gauge dependence is not affected, which is left in the first term on the second line of Eq.~(\ref{eq:WdecayM1}). Therefore, we can observe that the procedure to determine the counterterm of the lepton flavor mixing matrix is consistent with the gauge-independence of the decay rate.
	
\section{Matching Relations}
	
	\label{sec:conversion}
	
Given the counterterms of physical masses and flavor mixing matrices both in the on-shell scheme and in the $\overline{\rm MS}$ scheme calculated in Ref.~\cite{Huang:2025ubs}, we may figure out the matching relations between these two sets of parameters. For a bare parameter $a^{}_0$, one can decompose it into the renormalized one and the corresponding counterterm in both the $\overline{\rm MS}$ scheme and the on-shell scheme, i.e.,
	\begin{eqnarray}
		a_0^{} = \overline{a}(\mu) + \overline{\delta a} = a + \delta a (\mu) \;,
	\end{eqnarray}
where the quantities in the $\overline{\rm MS}$ scheme are overlined, and the dependence on the 't Hooft mass scale $\mu$ in the dimensional regularization is labeled explicitly. Since the $\overline{\rm MS}$ counterterms contain only the UV-divergent terms proportional to $\Delta$ at the one-loop level, which are the same as those divergences in the counterterms in the on-shell scheme, the matching relations between the on-shell parameters and their $\overline{\rm MS}$ counterparts are established through
	\begin{eqnarray}
		\overline{a}(\mu) = a + \left.\delta a (\mu)\right|_{\rm finite}  \;,
	\end{eqnarray}
where the subscript ``finite" implies that only the UV-finite terms are retained. Therefore, with the counterterms determined by the on-shell renormalization conditions, one can get the matching relations for all the physical parameters. 

First of all, for the running fine-structure constant $\overline{\alpha}(\mu)$, or equivalently the electromagnetic gauge coupling $\overline{e}(\mu)$, at the energy scale $\mu = m^{}_Z$ is related to the on-shell parameters by
	\begin{eqnarray}
		\overline{e}(m_Z^{}) = e\left\{1 - \frac{e^2}{16\pi^2} \left[\frac{7}{2}  \ln\left(\frac{m_Z^2}{m_W^2}\right) + \frac{1}{3} - \frac{2}{3} \sum_{f=q,l} N_{\rm c}^f Q_f^2 \ln\left(\frac{m_Z^2}{m_f^2}\right) \right] \right\} \;,
	\end{eqnarray}
	where $Q_f^{}$ is the corresponding electric charge of the fermion $f$. Here we have simply set $\mu=m_Z^{}$ since there exist precise measurements of the running fine-structure constant at the $Z$-pole. 
	
	Then, for the masses of elementary particles, we give the running masses at the energy scale of their on-shell masses, which are identical to their pole masses at the one-loop level~\cite{Sirlin:1991fd,Stuart:1991xk,Sirlin:1991rt}. For example, the running mass of the $W$-boson at $\mu=m_W^{}$ can be expressed in terms of the whole set of on-shell parameters as
	\begin{eqnarray}
		&& \overline{m}_W^2(m_W^{}) \nonumber \\
		&=& m_W^2 - \frac{\alpha}{144 \pi s^2 m_h^2 m_W^2} \left[ m_h^2 \left(628 m_W^4-69 m_W^2 m_Z^2-3 m_Z^4\right) \right. \nonumber \\
		&& \left. +45 m_h^4 m_W^2-3 m_h^6+18 m_W^2 \left(2 m_W^4+m_Z^4\right)  \right]  \nonumber \\
		&& - \frac{\alpha}{48 \pi s^2 m_W^2} \left(m_h^4-4 m_h^2 m_W^2+12 m_W^4\right) \Lambda\left(m_W^2;m_h^{},m_W^{}\right) \nonumber \\
		&& + \frac{\alpha}{48 \pi s^2 m_W^2 m_Z^2} \left(68 m_W^4 m_Z^2-16 m_W^2 m_Z^4+48 m_W^6-m_Z^6\right)  \Lambda\left(m_W^2;m_W^{},m_Z^{}\right) \nonumber \\
		&& + \frac{\alpha}{96 \pi s^2 m_h^2 m_W^4} \left[m_h^2 \left(84 m_W^4 m_Z^2-14 m_W^2 m_Z^4 -m_Z^6\right)+36 m_W^4 m_Z^4\right] \ln \left(\frac{m_W^2}{m_Z^2}\right) \nonumber \\
		&& - \frac{\alpha}{96 \pi s^2 m_W^4} m_h^4 \left(m_h^2-6 m_W^2\right) \ln \left(\frac{m_W^2}{m_h^2}\right) \nonumber \\
		&&  - \frac{\alpha}{48 \pi s^2} \sum_{\{ l_\alpha^{},\chi_i^{}\}} \left|{\cal B}_{\alpha i}^{}\right|^2 \left\{ 4 \left(\widehat{m}_i^2+m_{\alpha}^2\right) + \frac{2 \left(\widehat{m}_i^2-m_{\alpha }^2\right)^2}{m_W^2} - \frac{20}{3} m_W^2 \right. \nonumber \\
		&&  + \frac{2}{m_W^2} \left(\widehat{m}_i^4+m_{\alpha }^4-2 \widehat{m}_i^2 m_{\alpha }^2+\widehat{m}_i^2 m_W^2+m_{\alpha }^2 m_W^2-2 m_W^4\right) \Lambda\left(m_W^2;m_{\alpha }^{},\widehat{m}_i^{}\right)   \nonumber \\
		&& \left. - \frac{\left(\widehat{m}_i^2-m_{\alpha}^2\right)^3 }{m_W^4} \ln \left(\frac{\widehat{m}_i^2}{m_{\alpha }^2}\right) + \left(3 \widehat{m}_i^2+3 m_{\alpha }^2-2 m_W^2\right) \left[\ln \left(\frac{m_W^2}{\widehat{m}_i^2}\right)+(i\to \alpha)\right]  \right\} \nonumber \\
		&&  - \frac{\alpha}{16 \pi s^2} \sum_{\{q,q'\}} \left|{\bf V}_{q q'}^{\rm CKM}\right|^2 \left\{ 4 \left(m_{q'}^2+m_q^2\right) + \frac{2 \left(m_q^2-m_{q'}^2\right)^2}{m_W^2} - \frac{20}{3} m_W^2 \right. \nonumber \\
		&& \left.  + \frac{2}{m_W^2} \left(m_q^4 + m_{q'}^4 - 2 m_q^2 m_{q'}^2 + m_q^2 m_W^2 + m_{q'}^2 m_W^2 - 2 m_W^4\right) \Lambda\left(m_W^2;m_q^{},m_{q'}^{}\right) \right. \nonumber \\
		&& \left. - \frac{\left(m_q^2-m_{q'}^2\right)^3}{m_W^4} \ln \left(\frac{m_q^2}{m_{q'}^2}\right) + \left(3 m_{q'}^2+3 m_q^2-2 m_W^2\right) \left[\ln \left(\frac{m_W^2}{m_{q}^2}\right)+(q\to q')\right] \right\} \nonumber \\
		&&  + \frac{\alpha}{2 \pi s^2} \sum_{f=q,l} N_{\rm c}^f \frac{m_f^4}{m_h^2} \left[\ln \left(\frac{m_W^2}{m_f^2}\right)+1\right] + \frac{\alpha}{2 \pi s^2} \sum_{k} {\cal C}_{kk}^{} \frac{\widehat{m}_k^4}{m_h^2} \left[\ln \left(\frac{m_W^2}{\widehat{m}_k^2}\right)+1\right] \;,
	\end{eqnarray}
where the last line comes from the tadpole contributions to the $W$-boson self-energy with fermions running in the loop, and the color factor has already been included in the terms with the CKM matrix elements. It is worth mentioning that the bosonic and quark contributions to the $W$-boson self-energy are the same as those in the SM, while the leptonic part should be modified by including the contribution from massive Majorana neutrinos and the lepton flavor mixing.
	
	Similarly, for the running mass of the $Z$-boson, we have
	\begin{eqnarray}
		&& \overline{m}_Z^2 (m_Z^{}) \nonumber \\
		&=& m_Z^2 + \frac{\alpha}{144 \pi s^2 m_h^2 m_W^2 m_Z^2} \left[m_h^2 \left(70 m_Z^6 + 70 m_Z^4 m_W^2 - 408 m_Z^2 m_W^4 - 288 m_W^6  \right) \right. \nonumber \\
		&& \left. -45 m_h^4 m_Z^4 + 3 m_h^6 m_Z^2-18 \left(2 m_W^4 m_Z^4+m_Z^8\right)\right] \nonumber \\
		&& +\frac{\alpha}{48 \pi s^2 m_W^2} \left(12 m_Z^4-4 m_h^2 m_Z^2+m_h^4\right) \Lambda\left(m_Z^2;m_h^{},m_Z^{}\right) \nonumber \\
		&& - \frac{\alpha}{48 \pi s^2 m_W^2 m_Z^2} \left(48 m_W^6+68 m_W^4 m_Z^2-16 m_W^2 m_Z^4-m_Z^6\right) \Lambda\left(m_Z^2;m_W^{},m_W^{}\right) \nonumber \\
		&& + \frac{\alpha}{48 \pi s^2 m_h^2 m_W^2} \left[36 m_W^4 m_Z^2-m_h^2 \left(m_Z^4+14 m_W^2 m_Z^2-84 m_W^4\right)\right] \ln \left(\frac{m_W^2}{m_h^2}\right) \nonumber \\
		&& + \frac{\alpha}{96 \pi s^2 m_h^2 m_W^2 m_Z^2} \left[m_h^8 - 6 m_h^6 m_Z^2  \right. \nonumber \\
		&& \left. +2 m_h^2 m_Z^2 \left(m_Z^4 + 14 m_W^2 m_Z^2-84 m_W^4\right) -72 m_W^4 m_Z^4\right] \ln \left(\frac{m_Z^2}{m_h^2}\right) \nonumber \\
		&&  + \frac{\alpha m_Z^2}{12 \pi s^2 m_W^2} \sum_{f=q,l} N_{\rm c}^f \left\{ \vphantom{\ln \left(\frac{m_Z^2}{m_f^2}\right)} \frac{5}{3} \left[\left(c_{\rm A}^{f}\right)^2+\left(c_{\rm V}^{f}\right)^2\right] m_Z^2 + 4 \left[\left(c_{\rm V}^{f}\right)^2 - 2 \left(c_{\rm A}^{f}\right)^2\right] m_f^2 \right. \nonumber \\
		&& + \left[\left(c_{\rm A}^{f}\right)^2 \left(m_Z^2-4 m_f^2\right)+\left(c_{\rm V}^{f}\right)^2 \left(2 m_f^2+m_Z^2\right)\right] \Lambda\left(m_Z^2;m_f^{},m_f^{}\right) \nonumber \\
		&& \left. +  \left[\left(c_{\rm A}^{f}\right)^2 \left(m_Z^2-6 m_f^2\right)+\left(c_{\rm V}^{f}\right)^2 m_Z^2\right] \ln \left(\frac{m_Z^2}{m_f^2}\right) \right\} \nonumber \\
		&& - \frac{\alpha}{96 \pi s^2 m_W^2} \sum_{i,j} \left\{ \vphantom{\frac{ \left|{\cal C}_{ij}^{}\right|^2}{m_Z^2}} \vphantom{\sqrt{1}} 12 \widehat{m}_i^{} \widehat{m}_j^{} m_Z^2 \left({\cal C}_{ij}^2+{\cal C}_{ij}^{*2}\right) \right. \nonumber \\
		&& +\frac{2}{3} \left|{\cal C}_{ij}^{}\right|^2 \left[2 m_Z^2 \left(3 \widehat{m}_i^2+3 \widehat{m}_j^2-5 m_Z^2\right)+3 \left(\widehat{m}_i^2-\widehat{m}_j^2\right)^2\right] \nonumber \\
		&& + 2 \left\{ \left|{\cal C}_{ij}^{}\right|^2 \left[m_Z^2 \left(\widehat{m}_i^2+\widehat{m}_j^2-2 m_Z^2\right)+\left(\widehat{m}_i^2-\widehat{m}_j^2\right)^2\right] \right. \nonumber \\
		&& \left. + 3 \widehat{m}_i^{} \widehat{m}_j^{} m_Z^2 \left({\cal C}_{ij}^2+{\cal C}_{ij}^{*2}\right) \vphantom{\left|{\cal C}_{ij}^{}\right|^2} \vphantom{\sqrt{1}}\right\} \Lambda\left(m_Z^2;\widehat{m}_i^{},\widehat{m}_j^{}\right) \nonumber \\
		&& +3 \widehat{m}_i^{} \widehat{m}_j^{} \left({\cal C}_{ij}^2+{\cal C}_{ij}^{*2}\right) \left\{m_Z^2 \left[\ln \left(\frac{m_Z^2}{\widehat{m}_i^2}\right)+(i\to j)\right]+\left(\widehat{m}_j^2-\widehat{m}_i^2\right) \ln \left(\frac{\widehat{m}_i^2}{\widehat{m}_j^2}\right)\right\} \nonumber \\
		&& \left. +\frac{ \left|{\cal C}_{ij}^{}\right|^2}{m_Z^2} \left\{m_Z^4 \left(3 \widehat{m}_i^2+3 \widehat{m}_j^2-2 m_Z^2\right) \left[\ln \left(\frac{m_Z^2}{\widehat{m}_i^2}\right)+(i\to j) \right] - \left(\widehat{m}_i^2-\widehat{m}_j^2\right)^3 \ln \left(\frac{\widehat{m}_i^2}{\widehat{m}_j^2}\right)\right\} \right\} \nonumber \\
		&&  + \frac{\alpha m_Z^2 }{2\pi s^2 m_W^2} \sum_{f=q,l} N_{\rm c}^f \frac{m_{f}^4}{m_h^2}  \left[\ln \left(\frac{m_Z^2}{m_{f}^2}\right)+1\right] +  \frac{\alpha m_Z^2 }{2 \pi s^2 m_W^2} \sum_{i} {\cal C}_{ii}^{} \frac{\widehat{m}_i^4}{m_h^2} \left[\ln \left(\frac{m_Z^2}{\widehat{m}_i^2}\right)+1\right] \;.
	\end{eqnarray}
The tadpole contribution to the $Z$-boson self-energy has been taken into consideration to ensure the gauge-independence of the physical mass. The vector-type and axial-vector-type couplings of the neutral-current interaction are defined as $c_{\rm V}^f \equiv I_3^f - 2 s^2 Q_f^{}$ and $c_{\rm A}^f \equiv I_3^f$, respectively, with $I_3^f$ being the third component of the weak isospin for the charged fermion $f$.
	
	For the Higgs boson, its running mass at the energy scale of its pole mass reads
	\begin{eqnarray}
		&& \overline{m}_h^2 (m_h^{}) \nonumber \\
		&=& m_h^2 + \frac{\alpha}{16 \pi s^2 m_W^2} \left[9 m_h^4- 5 m_h^2 \left(2 m_W^2+ m_Z^2\right) + 6 \left(2 m_W^4+m_Z^4\right)\right] - \frac{3 \sqrt{3} \alpha m_h^4}{32 s^2 m_W^2} \nonumber \\
		&& + \frac{\alpha}{32 \pi s^2 m_W^2} \left(m_h^4-4 m_h^2 m_Z^2+12 m_Z^4\right) \Lambda\left(m_h^2;m_Z^{},m_Z^{}\right)  \nonumber \\
		&& + \frac{\alpha}{16 \pi s^2 m_W^2} \left(m_h^4-4 m_h^2 m_W^2+12 m_W^4\right) \Lambda\left(m_h^2;m_W^{},m_W^{}\right) \nonumber \\
		&& - \frac{\alpha m_h^2}{32 \pi s^2 m_W^2} \left(m_h^2-6 m_Z^2\right) \ln \left(\frac{m_Z^2}{m_h^2}\right) - \frac{\alpha m_h^2 }{16 \pi s^2 m_W^2} \left(m_h^2 - 6 m_W^2 \right) \ln \left(\frac{m_W^2}{m_h^2}\right) \nonumber \\
		&& - \frac{\alpha}{8 \pi s^2 } \sum_{f=q,l} N_{\rm c}^f \frac{m_f^2}{m_W^2} \left[10 m_f^2-2 m_h^2 + \left(4 m_f^2-m_h^2\right) \Lambda\left(m_h^2;m_f^{},m_f^{}\right)+\left(6 m_f^2-m_h^2\right) \ln \left(\frac{m_h^2}{m_f^2}\right)\right] \nonumber \\
		&& + \frac{\alpha}{32 \pi s^2 m_h^2 m_W^2} \sum_{i,j} \left\{ \vphantom{\left(\frac{1}{1}\right)^2}\widehat{m}_i^{} \widehat{m}_j^{} \left({\cal C}_{ij}^2 + {\cal C}_{ij}^{*2}\right) \left\{ \vphantom{\frac{1}{1}} 4 m_h^4 - 10 m_h^2 \left(\widehat{m}_i^2+\widehat{m}_j^2\right) \right. \right. \nonumber \\
		&& + 2 \left[m_h^4-2 m_h^2 \left(\widehat{m}_i^2+\widehat{m}_j^2\right)\right] \Lambda\left(m_h^2;\widehat{m}_i^{},\widehat{m}_j^{}\right)  \nonumber \\
		&& \left. +m_h^2 \left[m_h^2-3 \left(\widehat{m}_i^2+\widehat{m}_j^2\right)\right] \left[\ln \left(\frac{m_h^2}{\widehat{m}_i^2}\right)+(i\to j) \right] +2 \left(\widehat{m}_i^4-\widehat{m}_j^4\right) \ln \left(\frac{\widehat{m}_i^2}{\widehat{m}_j^2}\right)\right\} \nonumber \\
		&& + \left|{\cal C}_{ij}^{}\right|^2 \left\{ \vphantom{\ln \left(\frac{\widehat{m}_i^2}{\widehat{m}_j^2}\right)} 4 m_h^4 \left(\widehat{m}_i^2+\widehat{m}_j^2\right)- 2 m_h^2 \left[14 \widehat{m}_i^2 \widehat{m}_j^2 + 3 \left(\widehat{m}_i^4+\widehat{m}_j^4\right)\right] \right. \nonumber \\
		&& + 2 m_h^2 \left[m_h^2 \left(\widehat{m}_i^2+\widehat{m}_j^2\right)-6 \widehat{m}_i^2 \widehat{m}_j^2-\widehat{m}_i^4-\widehat{m}_j^4\right] \Lambda\left(m_h^2;\widehat{m}_i^{},\widehat{m}_j^{}\right) \nonumber \\
		&& +m_h^2 \left[m_h^2 \left(\widehat{m}_i^2+\widehat{m}_j^2\right)-2 \left(\widehat{m}_i^4 + 4 \widehat{m}_i^2 \widehat{m}_j^2+\widehat{m}_j^4\right)\right] \left[\ln \left(\frac{m_h^2}{\widehat{m}_i^2}\right)+(i\to j) \right] \nonumber \\
		&& \left.\left. +\left(5 \widehat{m}_i^4 \widehat{m}_j^2-5 \widehat{m}_i^2 \widehat{m}_j^4+\widehat{m}_i^6-\widehat{m}_j^6\right) \ln \left(\frac{\widehat{m}_i^2}{\widehat{m}_j^2}\right)\right\} \vphantom{\left(\frac{1}{1}\right)^2} \right\} \nonumber \\
		&& + \frac{3 \alpha}{4\pi s^2 }  \sum_{f=q,l} N_{\rm c}^f \frac{m_{f}^4}{m_W^2} \left[\ln \left(\frac{m_h^2}{m_{f}^2}\right)+1\right] + \frac{3 \alpha}{4\pi s^2 }  \sum_i {\cal C}_{ii}^{} \frac{\widehat{m}_{i}^4}{m_W^2} \left[\ln \left(\frac{\widehat{m}_h^2}{\widehat{m}_{i}^2}\right)+1\right]  \;,
	\end{eqnarray}
where the identity $\Lambda\left(m_h^2;m_h^{},m_h^{}\right) = - \pi/\sqrt{3}$ has been used.
	
For the running masses of quarks, we have
	\begin{eqnarray}\label{eq:runmq}
		&& \overline{m}_q^{}(m_q^{}) \nonumber \\
		&=& m_q^{} - \frac{\alpha}{288 \pi s^2 m_q^{} m_h^2 m_W^2} \left\{ \vphantom{\sqrt{1}} 36 m_h^4 m_q^2 + m_h^2 \left\{m_q^2 \left[m_W^2 \left(9-48 Q_q^{}\right)+48 m_Z^2 Q_q^{}\right] -54 m_q^4 \right. \right. \nonumber \\
		&& \left.\left. -12 m_W^2 m_Z^2 Q_q \left(6 Q_q+1\right) +18 m_W^4 \left(4 Q_q^2+1\right)+m_Z^4 \left(36 Q_q^2+1\right) \right\} +18 m_q^2 \left(2 m_W^4+m_Z^4\right) \vphantom{\sqrt{1}}\right\} \nonumber \\
		&& - \frac{\alpha m_Z^2}{64 \pi s^2 m_q^{} m_W^2} \left\{\left[8 \left(c_{\rm V}^q\right)^2-4\right] m_q^2+\left[4 \left(c_{\rm V}^q\right)^2+1\right] m_Z^2\right\} \Lambda\left(m_q^2;m_q^{},m_Z^{}\right) \nonumber \\
		&& + \frac{\alpha m_q^{}}{32 \pi s^2 m_W^2} \left(4 m_q^2-m_h^2\right) \Lambda\left(m_q^2;m_q^{},m_h^{}\right) \nonumber \\
		&& - \frac{\alpha m_Z^4}{128 \pi s^2  m_q^3 m_h^2 m_W^2}  \left\{ 24 m_q^4-6 m_q^2 m_h^2 + \left[4 \left(c_{\rm V}^q\right)^2+1\right] m_h^2 m_Z^2\right\} \ln \left(\frac{m_q^2}{m_Z^2}\right) \nonumber \\
		&& - \frac{3 \alpha m_q^{} m_W^2}{8 \pi s^2 m_h^2} \ln \left(\frac{m_q^2}{m_W^2}\right) - \frac{\alpha m_h^4}{64 \pi s^2 m_q^{} m_W^2} \ln \left(\frac{m_q^2}{m_h^2}\right) \nonumber \\
		&& + \frac{\alpha}{64 \pi s^2 m_q^3 m_W^2} \sum_{q'} \left|{\bf V}_{q q'}^{\rm CKM}\right|^2 \left\{ \vphantom{\ln \left(\frac{m_f^2}{m_{q'}^2}\right)} 2 m_q^2 \left(2 m_q^4+m_{q'}^4 - 5 m_q^2 m_{q'}^2+m_{q'}^2 m_W^2 \right) \right. \nonumber \\
		&& + 2 m_q^2 \left[ \left(m_q^2 - m_{q'}^2 \right)^2 + \left(m_q^2 + m_{q'}^2 \right) m_W^2 - 2 m_W^4\right] \Lambda\left(m_q^2;m_{q'}^{},m_W^{}\right)\nonumber \\
		&& + \left[3 m_q^2 m_{q'}^2 \left(m_{q'}^2 - m_q^2\right) + 3 m_W^4 \left(m_q^2 + m_{q'}^2\right) + m_q^6-m_{q'}^6-2 m_W^6\right] \ln \left(\frac{m_{q'}^2}{m_W^2}\right)   \nonumber \\
		&& \left. +2 m_q^4 \left(m_q^2-3 m_{q'}^2\right) \ln \left(\frac{m_q^2}{m_{q'}^2}\right)\right\} \nonumber \\
		&& + \frac{\alpha m_q^{}}{4\pi s^2 m_W^2} \sum_{f=q,l} N_{\rm c}^f \frac{m_{f}^4}{m_h^2} \left[\ln \left(\frac{m_q^2}{m_{f}^2}\right)+1\right]+ \frac{\alpha m_q^{}}{4\pi s^2 m_W^2} \sum_{i} {\cal C}_{ii}^{} \frac{\widehat{m}_{i}^4}{m_h^2} \left[\ln \left(\frac{m_q^2}{\widehat{m}_{i}^2}\right)+1\right] \;.
	\end{eqnarray}
Notice that the running masses of quarks are extracted in very different ways for light and heavy quarks~\cite{ParticleDataGroup:2024cfk, Chetyrkin:2000yt, Xing:2007fb, Xing:2011aa, Herren:2017osy, Martin:2019lqd, Huang:2020hdv}. For charged leptons, one can simply obtain their running masses $\overline{m}_\alpha^{}(m_\alpha^{})$ by performing the replacements $\left\{m_q^{}, m_{q'}^{}, Q_q^{}, c_{\rm V}^q, {\bf V}_{q q'}^{\rm CKM} \right\} \to \left\{m_\alpha^{}, \widehat{m}_i^{}, Q_l^{}, c_{\rm V}^l, {\cal B}_{\alpha i}^{} \right\}$ in Eq.~(\ref{eq:runmq}). 
	
	For Majorana neutrinos, we obtain
	\begin{eqnarray}
		&& \overline{\widehat{m}}_i^{}(\widehat{m}_i^{}) \nonumber \\
		&=&  \widehat{m}_i^{} - \frac{\alpha}{64 \pi s^2 m_h^2 m_W^2 \widehat{m}_i^{}} \left\{ \vphantom{\frac{1}{2}} 2 {\cal C}_{ii}^{} \left\{ 7 m_h^4 \widehat{m}_i^2 +4 \widehat{m}_i^2 \left(2 m_W^4+m_Z^4\right) \right. \right. \nonumber \\
		&& \left. + m_h^2 \left[\widehat{m}_i^2 \left(2 m_W^2+m_Z^2\right)-8 \widehat{m}_i^4+2 \left(2 m_W^4+m_Z^4\right)\right] \right\} \nonumber \\ 
		&& + 2 \sum_k \widehat{m}_k^2 \left[ \left|{\cal C}_{ik}^{}\right|^2 \left(m_h^4 -2  \widehat{m}_k^2 m_h^2 -  m_Z^2  m_h^2-6  \widehat{m}_i^2  m_h^2\right) - 2 \widehat{m}_i^{} \widehat{m}_k^{} m_h^2 \left({\cal C}_{ik}^2 + {\cal C}_{ki}^2 \right)\right]  \nonumber \\
		&& \left. - 4 m_h^2 \sum_\alpha \left|{\cal B}_{\alpha i}^{}\right|^2 \left(m_{\alpha }^4 +  m_W^2 m_{\alpha }^2 -5  m_{\alpha }^2 \widehat{m}_i^2 \right) \vphantom{\frac{1}{2}} \right\} \nonumber \\
		&& - \frac{\alpha}{16 \pi s^2 m_W^2 \widehat{m}_i^{}} \sum_\alpha  \left|{\cal B}_{\alpha i}^{}\right|^2 \left[2 m_W^4-\left(m_{\alpha }^2+\widehat{m}_i^2 \right)m_W^2-m_{\alpha }^4-\widehat{m}_i^4+2 m_{\alpha }^2 \widehat{m}_i^2\right]  \Lambda\left(\widehat{m}_i^2;m_W^{},m_{\alpha}^{}\right)   \nonumber \\
		&& + \frac{\alpha}{32 \pi s^2 m_W^2 \widehat{m}_i^{}}  \sum_k \left\{ \left|{\cal C}_{ik}^{}\right|^2 \left[ \left(\widehat{m}_i^2 - \widehat{m}_k^2 \right)^2 + \left(\widehat{m}_i^2 + \widehat{m}_k^2 \right) m_Z^2- 2   m_Z^4 \right] \right. \nonumber \\
		&& \left. + 3 \widehat{m}_i^{} \widehat{m}_k^{} m_Z^2 \left({\cal C}_{ik}^2 + {\cal C}_{ki}^2 \right) \vphantom{\sqrt{1}}  \right\} \Lambda\left(\widehat{m}_i^2;m_Z^{},\widehat{m}_k^{}\right) \nonumber \\
		&& + \frac{\alpha}{32 \pi s^2 m_W^2 \widehat{m}_i^{}}  \sum_k \left\{ \vphantom{\sqrt{1}} \left|{\cal C}_{ik}^{}\right|^2 \left[ \widehat{m}_i^4 + \widehat{m}_k^4 +6  \widehat{m}_k^2 \widehat{m}_i^2 - m_h^2 \left(\widehat{m}_i^2 + \widehat{m}_k^2\right) \right] \right. \nonumber \\
		&& \left. + \widehat{m}_k^{} \widehat{m}_i^{} \left({\cal C}_{ik}^2 + {\cal C}_{ki}^2\right) \left(2 \widehat{m}_i^2+2  \widehat{m}_k^2 - m_h^2 \right)  \vphantom{\sqrt{1}}\right\} \Lambda\left(\widehat{m}_i^2;m_h^{},\widehat{m}_k^{}\right) \nonumber \\
		&& + \frac{\alpha}{32 \pi s^2 m_W^2 \widehat{m}_i^3} \sum_\alpha \left|{\cal B}_{\alpha i}^{}\right|^2 \left\{ \widehat{m}_i^6 \left[\ln \left(\frac{m_h^2}{m_W^2}\right)+(W\to\alpha) \right] -3 m_{\alpha }^2 \widehat{m}_i^4 \left[\ln \left(\frac{\widehat{m}_i^2}{m_W^2}\right)+(W\to\alpha) \right] \right. \nonumber \\
		&& \left. +  \left(2 m_W^6 + m_{\alpha }^6 - 3 m_{\alpha }^2 m_W^4  - 3 \widehat{m}_i^2 m_W^4 - 3 \widehat{m}_i^2 m_{\alpha }^4 \right) \ln \left(\frac{m_W^2}{m_{\alpha }^2}\right) \right\} \nonumber \\
		&& - \frac{\alpha {\cal C}_{ii}^{} \widehat{m}_i^{}}{16 \pi s^2 m_h^2 m_W^2}  \left\{ 12 m_W^4 \ln\left(\frac{m_h^2}{m_W^2}\right) + 6 m_Z^4 \ln\left(\frac{m_h^2}{m_Z^2}\right) \right. \nonumber \\
		&& \left. + \left[3 m_h^4-2 \widehat{m}_i^2 m_h^2+6 \left(2 m_W^4+m_Z^4\right)\right] \ln \left(\frac{\widehat{m}_i^2}{m_h^2}\right) \right\} \nonumber \\
		&& + \frac{\alpha}{64 \pi s^2 m_W^2 \widehat{m}_i^2} \sum_k \left({\cal C}_{ik}^2+{\cal C}_{ki}^2\right) \widehat{m}_k^{} \left\{3 m_Z^2 \left(m_Z^2+\widehat{m}_i^2-\widehat{m}_k^2\right)  \ln\left(\frac{\widehat{m}_k^2}{m_Z^2}\right) \right. \nonumber \\
		&& \left. + \left[m_h^4-3 \left(\widehat{m}_i^2+\widehat{m}_k^2\right) m_h^2+2 \left(\widehat{m}_i^4+3 m_Z^2 \widehat{m}_i^2+\widehat{m}_k^4\right)\right] \ln\left(\frac{m_h^2}{\widehat{m}_k^2}\right)  \right\}  \nonumber \\
		&& + \frac{\alpha}{64 \pi s^2 m_W^2 \widehat{m}_i^3} \sum_k \left|{\cal C}_{ik}^{}\right|^2 \left\{ 4 \widehat{m}_k^2 \widehat{m}_i^4 \ln\left(\frac{\widehat{m}_i^2}{m_h^2}\right) + \left[3 \widehat{m}_i^6+\widehat{m}_k^6 + 5 \widehat{m}_k^4 \widehat{m}_i^2-\widehat{m}_i^4\widehat{m}_k^2 \right.\right. \nonumber \\
		&& \left. +m_h^4 \left(\widehat{m}_i^2+\widehat{m}_k^2\right)-2 m_h^2 \left(\widehat{m}_i^4+4 \widehat{m}_k^2 \widehat{m}_i^2+\widehat{m}_k^4\right)\right] \ln\left(\frac{m_h^2}{\widehat{m}_k^2}\right) \nonumber \\
		&& \left. + \left[\widehat{m}_i^6-\widehat{m}_k^6-2 m_Z^6+3 \left(\widehat{m}_k^2 + \widehat{m}_i^2 \right) m_Z^4+ 3 \left(\widehat{m}_i^2 \widehat{m}_k^4- \widehat{m}_i^4 \widehat{m}_k^2\right)\right] \ln\left(\frac{\widehat{m}_k^2}{m_Z^2}\right)  \right\} \nonumber \\
		&& + \frac{\alpha {\cal C}_{ii}^{} \widehat{m}_i^{}}{2\pi s^2 m_W^2}  \sum_{f=q,l} N_{\rm c}^f \frac{m_f^4}{m_h^2} \left[\ln \left(\frac{\widehat{m}_i^2}{m_f^2}\right)+1\right] + \frac{\alpha {\cal C}_{ii}^{} \widehat{m}_i^{}}{2 \pi s^2 m_W^2} \sum_k {\cal C}_{kk}^{} \frac{\widehat{m}_k^4}{m_h^2} \left[\ln \left(\frac{\widehat{m}_i^2}{\widehat{m}_k^2}\right)+1\right]  \;.
	\end{eqnarray}
	
	Finally, for completeness, we should also provide the matching relations of the flavor mixing matrices. We choose to fix the energy scale at the on-shell mass of the $Z$-boson. For the CKM matrix, we find the relation as follows
	\begin{eqnarray}
		&& \overline{{\bf V}^{\rm CKM}_{ud}} (m_Z^{}) \nonumber \\
		&=& {\bf V}^{\rm CKM}_{ud} + \frac{\alpha}{64 \pi s^2 m_W^2} \sum_{u'\neq u} \sum_{d'} {\bf V}_{u d'}^{\rm CKM} {\bf V}_{u' d}^{\rm CKM} {\bf V}_{u' d'}^{\rm CKM*} \nonumber \\
		&& \times \left\{ \vphantom{\left(\frac{1}{1}\right)^2} \frac{1}{\left(m_u^2-m_{u'}^2\right)} \left\{ \vphantom{\frac{1}{2}} 2 m_{d'}^2 \left(2m_{d'}^2 + 2m_W^2 -5m_{u'}^2-5 m_u^2 \right) \right. \right. \nonumber \\
		&& + \left\{ 2 \left[m_u^4 + m_{d'}^4 +\left(m_u^2 + m_{d'}^2 \right) m_W^2 - 2 m_u^2 m_{d'}^2 - 2 m_W^4 \right] \Lambda\left(m_u^2;m_W^{},m_{d'}^{}\right) + \left( u \to u'\right)  \right\}  \nonumber \\
		&& - \frac{1}{m_u^2 m_{u'}^2} \left[ m_{d'}^6 \left(m_{u'}^2+m_u^2\right) - 6 m_u^2 m_{d'}^4 m_{u'}^2 + 3 m_{d'}^2 \left(m_{u'}^2+m_u^2\right) \left(m_u^2 m_{u'}^2-m_W^4\right) \right. \nonumber \\
		&& \left. -m_u^6 m_{u'}^2 - m_u^2 m_{u'}^6 -  6 m_W^4 m_u^2 m_{u'}^2  + 2 m_W^6 \left(m_u^2 +  m_{u'}^2\right)\right] \ln \left(\frac{m_{d'}^2}{m_W^2}\right) \nonumber \\
		&& \left. +2 \left[m_{u'}^4+m_u^4-3 m_{d'}^2 \left(m_{u'}^2+m_u^2\right)\right] \ln \left(\frac{m_Z^2}{m_{d'}^2}\right)\right\} \nonumber \\
		&& \left. + \left(u\leftrightarrow d, u' \leftrightarrow d' \right) \vphantom{\left(\frac{1}{1}\right)^2} \right\} \;.
	\end{eqnarray}
In a similar way, we derive the relation for the lepton flavor mixing matrix
	\begin{eqnarray}
		&& \overline{{\cal B}_{\alpha i}^{}}(m_Z^{}) \nonumber \\
		&=& {\cal B}_{\alpha i}^{} + \frac{\alpha}{64 \pi s^2 m_W^2} \sum_{\beta\neq\alpha} \sum_k \frac{{\cal B}_{\alpha k}^{} {\cal B}_{\beta i}^{} {\cal B}_{\beta k}^* }{\left(m_{\alpha }^2-m_{\beta }^2\right)} \left\{ \vphantom{\frac{1}{2}} 2 \widehat{m}_k^2 \left(2 \widehat{m}_k^2+2 m_W^2-5 m_{\alpha }^2-5 m_{\beta }^2\right) \right. \nonumber \\
		&&  +\left\{2 \left[\widehat{m}_k^4+m_{\alpha}^4+ \left(\widehat{m}_k^2 + m_\alpha^2\right) m_W^2-2 \widehat{m}_k^2 m_{\alpha }^2 -2 m_W^4\right] \Lambda\left(m_{\alpha }^2;m_W^{},\widehat{m}_k^{}\right) + (\alpha\to\beta)\right\} \nonumber \\
		&& -\frac{1}{m_{\alpha }^2 m_{\beta }^2}  \left[ \widehat{m}_k^6 \left(m_{\alpha }^2+m_{\beta }^2\right)-6 \widehat{m}_k^4 m_{\alpha }^2 m_{\beta }^2+3 \widehat{m}_k^2 \left(m_{\alpha }^2+m_{\beta }^2\right) \left(m_{\alpha }^2 m_{\beta }^2 - m_W^4\right) \right. \nonumber \\
		&& \left. - m_{\alpha }^6 m_{\beta }^2 - m_{\alpha }^2 m_{\beta }^6 -6 m_{\alpha }^2 m_{\beta }^2 m_W^4 +2 m_W^6 \left(m_{\alpha }^2+m_{\beta }^2\right)\right] \ln \left(\frac{\widehat{m}_k^2}{m_W^2}\right) \nonumber \\
		&& \left.  + 2  \left[m_{\alpha}^4+m_{\beta}^4-3 \widehat{m}_k^2 \left(m_{\alpha}^2+m_{\beta}^2\right)\right] \ln \left(\frac{m_Z^2}{\widehat{m}_k^2}\right) \right\} \nonumber \\
		&& + \frac{\alpha}{4\pi s^2}  \sum_{j\neq i} \frac{{\cal B}_{\alpha j}^{}}{\widehat{m}_i^2-\widehat{m}_j^2}  \left\{\sum_{\rho} \frac{ {\cal B}_{\rho i}^{} {\cal B}_{\rho j}^{*}  m_{\rho}^2}{8 m_W^2 }  \left(2m_W^2+2m_{\rho }^2-5 \widehat{m}_i^2-5 \widehat{m}_j^2\right) \right.  \nonumber \\
		&& + \sum_\rho \frac{ {\cal B}_{\rho i}^{*} {\cal B}_{\rho j}^{}  m_{\rho }^2 }{8 m_W^2 \widehat{m}_i^{} \widehat{m}_j^{}} \left[\left(m_W^2+m_{\rho }^2\right) \left(\widehat{m}_i^2+\widehat{m}_j^2\right) -10 \widehat{m}_i^2  \widehat{m}_j^2  \right] \nonumber \\
		&& + \sum_k \frac{\widehat{m}_k^3}{2 m_W^2 }\left( \widehat{m}_i^{} {\cal C}_{ik}^{} {\cal C}_{jk}^{} + \widehat{m}_j^{} {\cal C}_{ki}^{} {\cal C}_{kj}^{} \right)+ \sum_k  \frac{ \widehat{m}_k^2 {\cal C}_{jk}^{} {\cal C}_{ki}^{} }{8 m_W^2 } \left(3 \widehat{m}_i^2+3 \widehat{m}_j^2+2 \widehat{m}_k^2-m_h^2+m_Z^2\right) \nonumber \\ 
		&& + \sum_k \frac{  {\cal C}_{ik}^{} {\cal C}_{kj}^{}  \widehat{m}_k^2}{16 m_W^2 \widehat{m}_i^{} \widehat{m}_j^{} } \left[\widehat{m}_i^2 \left(m_Z^2+12 \widehat{m}_j^2+2 \widehat{m}_k^2\right)+\widehat{m}_j^2 \left(m_Z^2+2 \widehat{m}_k^2\right)-m_h^2 \left(\widehat{m}_i^2+\widehat{m}_j^2\right)\right] \nonumber \\
		&& -\frac{1}{16 m_h^2 m_W^2 }  {\cal C}_{ji}^{} \left\{ \vphantom{\sqrt{1}} 7 \left(\widehat{m}_i^2+\widehat{m}_j^2\right) m_h^4 +4 \left(2 m_W^4+m_Z^4\right) \left(\widehat{m}_i^2+\widehat{m}_j^2\right) \right. \nonumber \\
		&& \left. - \left[6 \widehat{m}_i^4+6 \widehat{m}_j^4-\left(2 m_W^2+m_Z^2-4 \widehat{m}_j^2\right) \widehat{m}_i^2-\left(2 m_W^2+m_Z^2\right) \widehat{m}_j^2-4 \left(2 m_W^4+m_Z^4\right)\right] m_h^2 \vphantom{\sqrt{1}} \right\} \nonumber \\
		&& -\frac{1}{8 m_h^2 m_W^2 \widehat{m}_i^{} \widehat{m}_j^{} }  {\cal C}_{ij}^{} \left\{ \vphantom{\sqrt{1}} 7 \widehat{m}_i^2 \widehat{m}_j^2 m_h^4 +4 \left(2 m_W^4+m_Z^4\right) \widehat{m}_i^2 \widehat{m}_j^2 \right. \nonumber \\
		&& \left. - \left\{4 \widehat{m}_j^2 \widehat{m}_i^4-\left[2 m_W^4+m_Z^4-4 \widehat{m}_j^4+\left(2 m_W^2+m_Z^2\right) \widehat{m}_j^2\right] \widehat{m}_i^2-\left(2 m_W^4+m_Z^4\right) \widehat{m}_j^2 \right\} m_h^2 \vphantom{\sqrt{1}}\right\} \nonumber \\
		&& + \frac{1}{8 m_W^2} \sum_\rho \left\{ \left({\cal B}_{\rho i}^{} {\cal B}_{\rho j}^{*} + {\cal B}_{\rho i}^{*} {\cal B}_{\rho j}^{} \frac{\widehat{m}_j^{}}{\widehat{m}_i^{}}\right)  \right. \nonumber \\
		&& \left. \times \left[\left(m_{\rho}^2 - \widehat{m}_i^2 \right)^2 -2 m_W^4+\left(m_{\rho }^2 +\widehat{m}_i^2 \right) m_W^2 \right] \Lambda\left(\widehat{m}_i^2;m_W^{},m_{\rho }^{}\right)  + \left(\widehat{m}_i^{} \leftrightarrow \widehat{m}_j^{} \right) \vphantom{\frac{1}{2}} \right\}\nonumber \\
		&& -\frac{1}{16 m_W^2 \widehat{m}_i^{} } \sum_k  \left\{ \vphantom{\sqrt{1}} {\cal C}_{ik}^{} {\cal C}_{kj}^{} \widehat{m}_j^{} \left[m_h^2 \left(\widehat{m}_i^2+\widehat{m}_k^2\right)-\widehat{m}_i^4-6 \widehat{m}_k^2 \widehat{m}_i^2-\widehat{m}_k^4\right] \right. \nonumber \\
		&& +2 {\cal C}_{ki}^{} {\cal C}_{kj}^{} \widehat{m}_i^{} \widehat{m}_j^{} \widehat{m}_k^{} \left[m_h^2-2 \left(\widehat{m}_i^2+\widehat{m}_k^2\right)\right] \nonumber \\
		&&  -{\cal C}_{ik}^{} {\cal C}_{jk}^{} \widehat{m}_k^{} \left[\widehat{m}_i^4+3 \left(\widehat{m}_j^2+\widehat{m}_k^2\right) \widehat{m}_i^2-m_h^2 \left(\widehat{m}_i^2+\widehat{m}_j^2\right) +\widehat{m}_j^2 \widehat{m}_k^2 \right] \nonumber \\
		&& \left. -{\cal C}_{jk}^{} {\cal C}_{ki}^{} \widehat{m}_i^{} \left[\widehat{m}_k^4+3 \left(\widehat{m}_i^2 + \widehat{m}_j^2\right) \widehat{m}_k^2-m_h^2 \left(\widehat{m}_j^2+\widehat{m}_k^2\right)+\widehat{m}_i^2 \widehat{m}_j^2 \right] \vphantom{\sqrt{1}} \right\} \Lambda\left(\widehat{m}_i^2;m_h^{},\widehat{m}_k^{}\right)  \nonumber \\
		&& -\frac{ 1 }{16 m_W^2 \widehat{m}_j^{}}  \sum_k \left\{ \vphantom{\sqrt{1}} {\cal C}_{ik}^{} {\cal C}_{kj}^{} \widehat{m}_i^{} \left[m_h^2 \left(\widehat{m}_j^2+\widehat{m}_k^2\right)-\widehat{m}_j^4-6 \widehat{m}_k^2 \widehat{m}_j^2-\widehat{m}_k^4 \right] \right. \nonumber \\
		&&  +2 {\cal C}_{ik}^{} {\cal C}_{jk}^{} \widehat{m}_i^{} \widehat{m}_j^{} \widehat{m}_k^{} \left[m_h^2-2 \left(\widehat{m}_j^2+\widehat{m}_k^2\right)\right] \nonumber \\
		&& -{\cal C}_{ki}^{} {\cal C}_{kj}^{} \widehat{m}_k^{} \left[\widehat{m}_j^4+3 \left(\widehat{m}_i^2+\widehat{m}_k^2\right) \widehat{m}_j^2-m_h^2 \left(\widehat{m}_i^2+\widehat{m}_j^2\right)+\widehat{m}_i^2 \widehat{m}_k^2 \right] \nonumber \\
		&& \left. -{\cal C}_{jk}^{} {\cal C}_{ki}^{} \widehat{m}_j^{} \left[\widehat{m}_k^4+3 \left(\widehat{m}_i^2 + \widehat{m}_j^2\right) \widehat{m}_k^2-m_h^2 \left(\widehat{m}_i^2+\widehat{m}_k^2\right)+\widehat{m}_i^2 \widehat{m}_j^2 \right] \vphantom{\sqrt{1}} \right\} \Lambda\left(\widehat{m}_j^2;m_h^{},\widehat{m}_k^{}\right) \nonumber \\
		&& +\frac{1}{16 m_W^2 \widehat{m}_i^{}}  \sum_k  \left\{\vphantom{\sqrt{1}}  6 {\cal C}_{ki}^{} {\cal C}_{kj}^{} \widehat{m}_i^{} \widehat{m}_j^{}   \widehat{m}_k^{} m_Z^2+ 6 {\cal C}_{ik}^{} {\cal C}_{jk}^{} \widehat{m}_i^2 \widehat{m}_k^{} m_Z^2 \right. \nonumber \\
		&& +{\cal C}_{ik}^{} {\cal C}_{kj}^{} \widehat{m}_j^{}  \left[\widehat{m}_i^4+\widehat{m}_k^4-2 m_Z^4+\widehat{m}_k^2 m_Z^2+\widehat{m}_i^2 \left(m_Z^2-2 \widehat{m}_k^2\right)\right]  \nonumber \\
		&& \left.  +{\cal C}_{ki}^{} {\cal C}_{jk}^{} \widehat{m}_i^{} \left[\widehat{m}_i^4+\widehat{m}_k^4-2 m_Z^4+\widehat{m}_k^2 m_Z^2+\widehat{m}_i^2 \left(m_Z^2-2 \widehat{m}_k^2\right)\right] \vphantom{\sqrt{1}}\right\} \Lambda\left(\widehat{m}_i^2;m_Z^{},\widehat{m}_k^{}\right)  \nonumber \\
		&& +\frac{1}{16 m_W^2 \widehat{m}_j^{} }  \sum_k \left\{ \vphantom{\sqrt{1}} 6 {\cal C}_{kj}^{} {\cal C}_{ki}^{} \widehat{m}_j^2 \widehat{m}_k^{} m_Z^2 + 6 {\cal C}_{jk}^{} {\cal C}_{ik}^{} \widehat{m}_i^{} \widehat{m}_j \widehat{m}_k m_Z^2 \right. \nonumber \\
		&& +{\cal C}_{jk}^{} {\cal C}_{ki}^{} \widehat{m}_j^{} \left[\widehat{m}_j^4+\widehat{m}_k^4 -2 m_Z^4+\widehat{m}_k^2 m_Z^2+\widehat{m}_j^2 \left(m_Z^2-2 \widehat{m}_k^2\right)\right]  \nonumber \\
		&& \left. +{\cal C}_{kj}^{} {\cal C}_{ik}^{} \widehat{m}_i^{} \left[\widehat{m}_j^4+\widehat{m}_k^4 -2 m_Z^4+\widehat{m}_k^2 m_Z^2+\widehat{m}_j^2 \left(m_Z^2-2 \widehat{m}_k^2\right)\right]  \vphantom{\sqrt{1}} \right\} \Lambda\left(\widehat{m}_j^2;m_Z^{},\widehat{m}_k^{}\right)  \nonumber \\
		&& +\frac{1}{16 m_W^2 \widehat{m}_i^3 \widehat{m}_j^3 } \sum_\rho \left\{ \vphantom{\frac{1}{2}} {\cal B}_{\rho i}^{*} {\cal B}_{\rho j}^{} \left[\left(2 m_W^6 -3 m_{\rho}^2 m_W^4\right) \left(\widehat{m}_i^4 + \widehat{m}_j^4\right) +m_{\rho}^6 \left(\widehat{m}_i^4+\widehat{m}_j^4\right) \right.\right. \nonumber \\
		&& \left. +\left(\widehat{m}_i^4 \widehat{m}_j^4-3 \widehat{m}_i^2 \widehat{m}_j^2 m_W^4\right) \left(\widehat{m}_i^2+\widehat{m}_j^2\right) -3m_{\rho}^2 \widehat{m}_i^2 \widehat{m}_j^2 \left(2 \widehat{m}_i^2 \widehat{m}_j^2+ m_{\rho }^2 \widehat{m}_j^2+ m_{\rho }^2 \widehat{m}_i^2 \right)\right]  \nonumber \\
		&& +{\cal B}_{\rho i}^{} {\cal B}_{\rho j}^{*} \left[\left(\widehat{m}_j^3 \widehat{m}_i^7+\widehat{m}_j^7 \widehat{m}_i^3\right)-3\left( m_{\rho}^2 \widehat{m}_j^3 \widehat{m}_i^5+ m_{\rho}^2 \widehat{m}_j^5 \widehat{m}_i^3\right)-6\left( m_W^4 \widehat{m}_j^3 \widehat{m}_i^3+ m_{\rho }^4 \widehat{m}_j^3 \widehat{m}_i^3\right) \right. \nonumber \\
		&& \left. \left. + \widehat{m}_i^{} \widehat{m}_j^{} \left(m_{\rho }^6 +2 m_W^6  -3 m_W^4 m_{\rho }^2 \right) \left(\widehat{m}_i^2+\widehat{m}_j^2\right) \right] \vphantom{\frac{1}{2}}\right\}  \ln\left(\frac{m_W^2}{m_{\rho}^2}\right) \nonumber \\
		&& -\frac{1}{8 m_h^2 m_W^2 } \sum_\rho \left\{ \vphantom{\frac{1}{2}}  12 \left[2 {\cal C}_{ij}^{} \widehat{m}_i^{} \widehat{m}_j^{} + {\cal C}_{ji}^{} \left(\widehat{m}_i^2+\widehat{m}_j^2\right)\right] m_W^4 -{\cal B}_{\rho i}^{*} {\cal B}_{\rho j}^{} m_h^2 \widehat{m}_i^{} \widehat{m}_j^{} \left(\widehat{m}_i^2+\widehat{m}_j^2-6 m_{\rho}^2\right) \right. \nonumber \\
		&& \left. -{\cal B}_{\rho i}^{} {\cal B}_{\rho j}^{*} m_h^2 \left(\widehat{m}_i^4+\widehat{m}_j^4-3 m_{\rho}^2 \widehat{m}_i^2-3 m_{\rho}^2 \widehat{m}_j^2\right) \vphantom{\frac{1}{2}} \right\} \ln\left(\frac{m_h^2}{m_W^2}\right) \nonumber \\
		&& -\frac{1}{16 m_h^2 m_W^2 \widehat{m}_i^{} \widehat{m}_j^{}}  \sum_k \left\{ \vphantom{\frac{1}{2}} m_h^2 \widehat{m}_k^2 \left[\left({\cal C}_{ki}^{} {\cal C}_{kj}^{} \widehat{m}_i^{} \widehat{m}_k^{}-{\cal C}_{ik}^{} {\cal C}_{jk}^{} \widehat{m}_j^{} \widehat{m}_k^{} \right) \left(\widehat{m}_i^2-\widehat{m}_j^2\right)  \right. \right. \nonumber \\
		&& \left. -4 {\cal C}_{ik}^{} {\cal C}_{kj}^{} \widehat{m}_i^2 \widehat{m}_j^2 -2 {\cal C}_{jk}^{} {\cal C}_{ki}^{} \widehat{m}_i^{} \widehat{m}_j^{} \left(\widehat{m}_i^2+\widehat{m}_j^2\right) \right]  \nonumber \\
		&& + 6 m_h^2 m_{\rho}^2 \widehat{m}_i^{} \widehat{m}_j^{} \left[2 {\cal B}_{\rho i}^{*} {\cal B}_{\rho j}^{}  \widehat{m}_i^{} \widehat{m}_j^{}+{\cal B}_{\rho i}^{} {\cal B}_{\rho j}^{*} \left(\widehat{m}_i^2+\widehat{m}_j^2\right)\right] \nonumber \\
		&&  + \widehat{m}_i^{} \widehat{m}_j^{}  {\cal C}_{ji}^{} \left[ 6  \left(4  m_W^4 +2  m_Z^4  +  m_h^4  \right) \left( \widehat{m}_i^2 + \widehat{m}_j^2 \right)-  m_h^2 \left(3  \widehat{m}_i^4 +3 \widehat{m}_j^4 +2  \widehat{m}_i^2 \widehat{m}_j^2 \right)  \right] \nonumber \\
		&& \left. + 4 {\cal C}_{ij}^{}  \widehat{m}_i^2 \widehat{m}_j^2 \left[3 m_h^4-\left(\widehat{m}_i^2+\widehat{m}_j^2\right) m_h^2+6 \left(2 m_W^4+m_Z^4\right)\right] \vphantom{\frac{1}{2}}\right\} \ln\left(\frac{m_Z^2}{m_h^2}\right) \nonumber \\
		&& -\frac{1}{32 m_W^2 \widehat{m}_i^3 \widehat{m}_j^3 } \sum_k \left\{ \vphantom{\frac{1}{2}} 6 \widehat{m}_i^{} \widehat{m}_j^{} \widehat{m}_k^{} m_Z^2 \left({\cal C}_{ik}^{} {\cal C}_{jk}^{} \widehat{m}_i^{}+{\cal C}_{ki}^{} {\cal C}_{kj}^{} \widehat{m}_j^{} \right) \right. \nonumber \\
		&& \times   \left[\left(2 \widehat{m}_j^2+\widehat{m}_k^2-m_Z^2\right) \widehat{m}_i^2+ \left(\widehat{m}_k^2-m_Z^2\right) \widehat{m}_j^2 \right]  \nonumber \\
		&& +{\cal C}_{jk}^{} {\cal C}_{ki}^{} \left[ \widehat{m}_j^3 \widehat{m}_i^3 \left( \widehat{m}_i^4+\widehat{m}_j^4\right) + 2 \widehat{m}_i^{}   \widehat{m}_j^{}  m_Z^4 \left(m_Z^2 \widehat{m}_i^2+m_Z^2 \widehat{m}_j^2 -3 \widehat{m}_i^2 \widehat{m}_j^2  \right) \right. \nonumber \\
		&& +\widehat{m}_i^{} \widehat{m}_j^{} \widehat{m}_k^{} \left( \widehat{m}_i^2 \widehat{m}_k^5 - \widehat{m}_i^4 \widehat{m}_j^2 \widehat{m}_k^{} -2 \widehat{m}_i^2 \widehat{m}_j^2 \widehat{m}_k^3 -\widehat{m}_i^2 \widehat{m}_j^4 \widehat{m}_k^{}  +\widehat{m}_j^2 \widehat{m}_k^5 \right) \nonumber \\
		&& \left. + \widehat{m}_i^{} \widehat{m}_j^{} \widehat{m}_k^{} m_Z^2 \left(8 \widehat{m}_i^2 \widehat{m}_j^2 \widehat{m}_k^{} -3 m_Z^2 \widehat{m}_i^2 \widehat{m}_k^{} -3 m_Z^2 \widehat{m}_j^2 \widehat{m}_k^{} \right) \right] \nonumber \\
		&& + {\cal C}_{ik}^{} {\cal C}_{kj}^{} \left[\left(2 m_Z^6 -3 \widehat{m}_k^2 m_Z^4 \right) \left(\widehat{m}_i^4 + \widehat{m}_j^4\right) +\left(4 \widehat{m}_i^2 \widehat{m}_j^2 \widehat{m}_k^2 m_Z^2 -3 \widehat{m}_i^2 \widehat{m}_j^2 m_Z^4\right) \left(\widehat{m}_i^2+ \widehat{m}_j^2\right) \right. \nonumber \\ 
		&& \left. \left. +\widehat{m}_i^4 \left( \widehat{m}_j^6+ \widehat{m}_k^6\right)+ \widehat{m}_j^4 \left( \widehat{m}_i^6+\widehat{m}_k^6  \right)- \widehat{m}_i^2 \widehat{m}_j^2 \widehat{m}_k^2 \left( \widehat{m}_j^2 \widehat{m}_k^2+\widehat{m}_i^2 \widehat{m}_k^2+2 \widehat{m}_i^2 \widehat{m}_j^2\right)\right] \vphantom{\frac{1}{2}} \right\} \ln\left(\frac{\widehat{m}_k^2}{m_Z^2}\right) \nonumber \\
		&& +\frac{1}{16 m_h^2 m_W^2 \widehat{m}_i^{} \widehat{m}_j^{} }  \sum_k \left\{ \vphantom{\frac{1}{2}} {\cal C}_{ij}^{} \widehat{m}_i^2 \widehat{m}_j^2 \left[m_h^2 \left(\widehat{m}_i^2+\widehat{m}_j^2\right)-24 m_Z^4\right] \right.\nonumber \\
		&& +{\cal C}_{ji}^{} \widehat{m}_i^{} \widehat{m}_j^{} \left[m_h^2 \left(\widehat{m}_i^4+\widehat{m}_j^4\right)-12 m_Z^4 \left(\widehat{m}_i^2+\widehat{m}_j^2\right)\right]   \nonumber \\
		&& +m_h^2 \widehat{m}_k^2 \left\{ {\cal C}_{ik}^{} {\cal C}_{kj}^{} \left[ \left(2 m_Z^2-4 \widehat{m}_j^2+\widehat{m}_k^2\right) \widehat{m}_i^2+ \left(2 m_Z^2+\widehat{m}_k^2\right) \widehat{m}_j^2 \right] \right. \nonumber \\
		&& \left. \left.  -2 {\cal C}_{jk}^{} {\cal C}_{ki}^{} \widehat{m}_i^{} \widehat{m}_j^{} \left(\widehat{m}_i^2+\widehat{m}_j^2-\widehat{m}_k^2 -2 m_Z^2\right)\right\} \vphantom{\frac{1}{2}} \right\} \ln \left(\frac{m_h^2}{m_Z^2}\right) \nonumber \\
		&& +\frac{ 1 }{32 m_W^2 \widehat{m}_i^3 \widehat{m}_j^3 } \sum_k \left\{ \vphantom{\frac{1}{1}} {\cal C}_{ki}^{} {\cal C}_{kj}^{} \widehat{m}_i^{} \widehat{m}_k^{} \left\{ \left(\widehat{m}_i^4+\widehat{m}_j^2 \widehat{m}_i^2+2 \widehat{m}_j^4\right) m_h^4 \right. \right. \nonumber \\
		&& -2 \left[\left(2 \widehat{m}_j^2+\widehat{m}_k^2\right) \widehat{m}_i^4+2 \widehat{m}_j^2 \left(2 \widehat{m}_j^2+\widehat{m}_k^2\right) \widehat{m}_i^2+3 \widehat{m}_j^4 \widehat{m}_k^2\right] m_h^2\nonumber \\
		&& \left. +4 \widehat{m}_j^4 \widehat{m}_k^4 + \widehat{m}_i^4 \left(7 \widehat{m}_j^4+\widehat{m}_k^4\right) + \widehat{m}_i^2 \left(\widehat{m}_j^6+3 \widehat{m}_k^4 \widehat{m}_j^2\right)\right\} \nonumber \\
		&&  +{\cal C}_{jk}^{} {\cal C}_{ki}^{} \widehat{m}_i^{} \widehat{m}_j^{} \left\{ \left(\widehat{m}_i^4+\widehat{m}_k^2 \widehat{m}_i^2+\widehat{m}_j^4+\widehat{m}_j^2 \widehat{m}_k^2\right) m_h^4 \right. \nonumber \\
		&& -2 \left(\widehat{m}_i^2+\widehat{m}_j^2\right) \left[\widehat{m}_k^4+2 \widehat{m}_j^2 \widehat{m}_k^2+\widehat{m}_i^2 \left(\widehat{m}_j^2+2 \widehat{m}_k^2\right)\right] m_h^2 +\widehat{m}_j^2 \widehat{m}_k^4 \left(3 \widehat{m}_j^2+\widehat{m}_k^2\right) \nonumber \\
		&& \left.  +\widehat{m}_i^2 \widehat{m}_k^2 \left(3 \widehat{m}_j^4-8 m_Z^2 \widehat{m}_j^2+\widehat{m}_k^4\right)+\widehat{m}_i^4 \left(2 \widehat{m}_j^4+3 \widehat{m}_k^2 \widehat{m}_j^2+3 \widehat{m}_k^4\right)\right\} \nonumber \\
		&&  +{\cal C}_{ik}^{} {\cal C}_{jk}^{} \widehat{m}_j^{}\widehat{m}_k^{} \left\{ \widehat{m}_j^2 \widehat{m}_i^6+\left(7 \widehat{m}_j^4+4 \widehat{m}_k^4\right) \widehat{m}_i^4+3 \widehat{m}_j^2 \widehat{m}_k^4 \widehat{m}_i^2+\widehat{m}_j^4 \widehat{m}_k^4 \right. \nonumber \\
		&& \left. +m_h^4 \left(2 \widehat{m}_i^4+\widehat{m}_j^2 \widehat{m}_i^2+\widehat{m}_j^4\right)-2 m_h^2 \left[\left(4 \widehat{m}_j^2+3 \widehat{m}_k^2\right) \widehat{m}_i^4+2 \widehat{m}_j^2 \left(\widehat{m}_j^2+\widehat{m}_k^2\right) \widehat{m}_i^2+\widehat{m}_j^4 \widehat{m}_k^2\right]\right\} \nonumber \\
		&& + {\cal C}_{ik}^{} {\cal C}_{kj}^{} \left\{ \widehat{m}_j^4 \widehat{m}_i^6+\left[\widehat{m}_j^6+6 \widehat{m}_k^2 \widehat{m}_j^4+\left(3 \widehat{m}_k^4-4 m_Z^2 \widehat{m}_k^2\right) \widehat{m}_j^2+\widehat{m}_k^6\right] \widehat{m}_i^4 \right. \nonumber \\
		&&  +\widehat{m}_j^4 \widehat{m}_k^2 \left(3 \widehat{m}_k^2-4 m_Z^2\right) \widehat{m}_i^2+\widehat{m}_j^4 \widehat{m}_k^6+m_h^4 \left[\left(\widehat{m}_j^2+\widehat{m}_k^2\right) \widehat{m}_i^4+\widehat{m}_j^4 \widehat{m}_i^2+\widehat{m}_j^4 \widehat{m}_k^2\right] \nonumber \\
		&& \left.\left. -2 m_h^2 \left[\left(2 \widehat{m}_j^4+4 \widehat{m}_k^2 \widehat{m}_j^2+\widehat{m}_k^4\right) \widehat{m}_i^4+4 \widehat{m}_j^4 \widehat{m}_k^2 \widehat{m}_i^2+\widehat{m}_j^4 \widehat{m}_k^4\right]\right\} \vphantom{\frac{1}{1}} \right\} \ln \left(\frac{m_h^2}{\widehat{m}_k^2}\right) \nonumber \\
		&& +  \frac{1}{m_h^2 m_W^2 } \left[2 {\cal C}_{ij}^{} \widehat{m}_i^{} \widehat{m}_j^{}+{\cal C}_{ji}^{} \left(\widehat{m}_i^2+\widehat{m}_j^2\right)\right]  \sum_{f=q,l} N_{\rm c}^f m_f^4 \left[\ln \left(\frac{m_Z^2}{m_f^2}\right)+1\right] \nonumber \\
		&& \left. +  \frac{1}{m_h^2 m_W^2 } \left[2 {\cal C}_{ij}^{} \widehat{m}_i^{} \widehat{m}_j^{}+{\cal C}_{ji}^{} \left(\widehat{m}_i^2+\widehat{m}_j^2\right)\right]   \sum_{k} {\cal C}_{kk}^{} \widehat{m}_k^4 \left[\ln \left(\frac{m_Z^2}{\widehat{m}_k^2}\right)+1\right] \right\}  \nonumber \\
		&& + \frac{\alpha}{4\pi s^2} \sum_k \frac{ \left({\cal C}_{ik}^2-{\cal C}_{ki}^2\right) \widehat{m}_k^{}}{32 \widehat{m}_i^3 m_W^2} \left\{ \vphantom{\left(\frac{1}{1}\right)} 4 \widehat{m}_i^2 \widehat{m}_k^2 - 2 \widehat{m}_i^2 \left(\widehat{m}_i^2-\widehat{m}_k^2\right) \Lambda\left(\widehat{m}_i^2;m_h^{},\widehat{m}_k^{}\right) \right. \nonumber \\
		&&  +6 \widehat{m}_i^2 m_Z^2 \Lambda\left(\widehat{m}_i^2;m_Z^{},\widehat{m}_k^{}\right) + 3 m_Z^2 \left(\widehat{m}_i^2-\widehat{m}_k^2+m_Z^2\right) \ln \left(\frac{\widehat{m}_k^2}{m_Z^2}\right)  \nonumber \\
		&& \left. +2 \widehat{m}_i^2 \widehat{m}_k^2 \ln \left(\frac{m_Z^2}{m_h^2}\right)+ \left[m_h^2 \left(\widehat{m}_i^2-\widehat{m}_k^2\right)+6 \widehat{m}_i^2 m_Z^2-\widehat{m}_i^4+\widehat{m}_k^4\right]\ln \left(\frac{m_h^2}{\widehat{m}_k^2}\right) \right\} \;.
	\end{eqnarray}
As we have mentioned, the diagonal self-energies of Majorana neutrinos indeed contribute to the counterterm of the lepton flavor mixing matrix and thus to this matching relation as well. 

The matching relations between on-shell and $\overline{\rm MS}$ parameters have already been extensively studied within the SM. Restricting to only the electroweak corrections, one can find the results at the two-loop order for the fine-structure constant~\cite{Fanchiotti:1992tu,Erler:1998sy,Degrassi:2003rw,Degrassi:2014sxa,Martin:2018yow}, the masses of gauge bosons~\cite{Jegerlehner:2001fb,Jegerlehner:2002em,Degrassi:2014sxa,Kniehl:2015nwa,Martin:2015lxa,Martin:2015rea}, the Higgs boson~\cite{Casas:1994qy,Bezrukov:2012sa,Degrassi:2012ry,Buttazzo:2013uya,Martin:2014cxa}, the top quark~\cite{Bohm:1986rj,Hempfling:1994ar,Jegerlehner:2002em,Jegerlehner:2003py,Jegerlehner:2003sp,Eiras:2005yt,Bezrukov:2012sa,Jegerlehner:2012kn,Faisst:2003px,Faisst:2004gn,Kataev:2022dua,Kniehl:2014yia,Kniehl:2015nwa,Martin:2016xsp}, charged leptons~\cite{Arason:1991ic,Gray:1990yh,Bekavac:2007tk} and light quarks~\cite{Hempfling:1994ar,Jegerlehner:2002em,Kniehl:2004hfa,Martin:2005ch,Kniehl:2015nwa}. The one-loop relation for the CKM matrix can also be found in Ref.~\cite{Kniehl:2000rb}. Although the matching conditions are derived in the present work only at the one-loop level, the contributions from massive Majorana neutrinos and leptonic flavor mixing are worked out explicitly for the first time. 
	
	\section{Summary}
	
	\label{sec:sum}
	
	In this paper, we proceed to carry out one-loop renormalization of the type-I seesaw model in the on-shell scheme, and provide the matching relations between physical parameters in the on-shell scheme and those in the $\overline{\rm MS}$ scheme. With the Feynman rules derived in our previous work~\cite{Huang:2025ubs}, we calculate the one-loop two-point functions of gauge bosons, the Higgs boson and fermions in the $R_\xi^{}$ gauge, where the fine-structure constant $\alpha$, the physical masses $\left\{m_W^{},m_Z^{},m_h^{},m_f^{}\right\}$ and mixing matrices $\left\{{\bf V}_{}^{\rm CKM}, {\bf V}, {\bf R}\right\}$ are chosen as input parameters. While the counterterms for masses and wave functions are determined through the on-shell conditions, those for the flavor mixing matrices need to be properly modified in order to guarantee the gauge independence of physical observables. Generalizing the approach for the CKM matrix, we propose a practical method to fix the counterterm for the lepton flavor mixing matrix ${\cal B} \equiv ({\bf V} ~ {\bf R})$ in the type-I seesaw model. Furthermore, the running $\overline{\rm MS}$ parameters at a given energy scale are explicitly expressed in terms of the on-shell parameters and the contributions from the massive Majorana neutrinos are taken into account.
	
Although one-loop renormalization of the type-I seesaw model has now been accomplished in both on-shell and $\overline{\rm MS}$ schemes, more theoretical works are required to fully realize its precision tests. First, one can perform one-loop calculations of the observables for specific SM processes in the on-shell scheme and compare theoretical predictions with experimental measurements. In contrast, it is very likely that the calculation of the same observable has been done in the SM at the two or even higher loop order. A tentative solution to this mismatch problem of precisions is to incorporate the most advanced results in the SM and only one-loop corrections in the seesaw model. Second, since most precision measurements are done at low energies, it is in fact more convenient to work in the $\overline{\rm MS}$ scheme, where one can integrate heavy particles out one by one and construct an effective theory at the corresponding energy scale until the characteristic scale for the physical processes of our interest is reached. The connection between physical parameters in two successive effective theories is provided by the renormalization-group equations. To thoroughly test the type-I seesaw model, one needs to recalculate the low-energy observables and make a global-fit analysis of the model parameters. Based on our results in the present and previous works~\cite{Zhang:2021tsq,Zhang:2021jdf,Wang:2023bdw,Huang:2025ubs}, it is interesting to further establish a systematic and feasible framework for future tests of the seesaw mechanism for neutrino mass generation with low-energy precision data, high-energy collider experiments and cosmological observations. 
	
	\section*{Acknowledgements}
	
	This work was supported in part by the National Natural Science Foundation of China under grant No.~12475113, by the CAS Project for Young Scientists in Basic Research (YSBR-099), and by the Scientific and Technological Innovation Program of IHEP under grant No.~E55457U2. Loop integrals in this work are calculated with the help of {\tt Package-X}~\cite{Patel:2015tea,Patel:2016fam}.


\begin{thebibliography}{99}
		%\cite{Minkowski:1977sc}
		\bibitem{Minkowski:1977sc}
		P.~Minkowski,
		``$\mu \to {\rm e} \gamma$ at a rate of one out of $10^{9}$ muon decays?,''
		Phys. Lett. B \textbf{67}, 421-428 (1977).
%		doi:10.1016/0370-2693(77)90435-X
		%5545 citations counted in INSPIRE as of 15 Sep 2025
		
		%\cite{Yanagida:1979as}
		\bibitem{Yanagida:1979as}
		T.~Yanagida,
		``Horizontal gauge symmetry and masses of neutrinos,''
		Conf. Proc. C \textbf{7902131}, 95-99 (1979)
		KEK-79-18-95.
		%2491 citations counted in INSPIRE as of 15 Sep 2025
		
		%\cite{Gell-Mann:1979vob}
		\bibitem{Gell-Mann:1979vob}
		M.~Gell-Mann, P.~Ramond and R.~Slansky,
		``Complex Spinors and Unified Theories,''
		Conf. Proc. C \textbf{790927}, 315-321 (1979)
		[arXiv:1306.4669 [hep-th]].
		%4207 citations counted in INSPIRE as of 15 Sep 2025
		
		%\cite{Glashow:1979nm}
		\bibitem{Glashow:1979nm}
		S.~L.~Glashow,
		``The Future of Elementary Particle Physics,''
		NATO Sci. Ser. B \textbf{61}, 687 (1980).
%		doi:10.1007/978-1-4684-7197-7{\_}15
		%878 citations counted in INSPIRE as of 15 Sep 2025
		
		%\cite{Mohapatra:1979ia}
		\bibitem{Mohapatra:1979ia}
		R.~N.~Mohapatra and G.~Senjanovi\'{c},
		``Neutrino Mass and Spontaneous Parity Nonconservation,''
		Phys. Rev. Lett. \textbf{44}, 912-915 (1980).
%		doi:10.1103/PhysRevLett.44.912
		%7098 citations counted in INSPIRE as of 15 Sep 2025
		
		%\cite{Huang:2025ubs}
		\bibitem{Huang:2025ubs}
		J.~Huang and S.~Zhou,
		``One-loop Renormalization of the Type-I Seesaw Model in the Modified Minimal-subtraction Scheme,''
		[arXiv:2507.21691 [hep-ph]].
		%1 citations counted in INSPIRE as of 26 Aug 2025
		
		
		%\cite{Bardeen:1978yd}
		\bibitem{Bardeen:1978yd}
		W.~A.~Bardeen, A.~J.~Buras, D.~W.~Duke and T.~Muta,
		``Deep-inelastic scattering beyond the leading order in asymptotically free gauge theories,''
		Phys. Rev. D \textbf{18}, 3998 (1978).
%		doi:10.1103/PhysRevD.18.3998
		%1854 citations counted in INSPIRE as of 11 Sep 2025
		
		%\cite{Weinberg:1980wa}
		\bibitem{Weinberg:1980wa}
		S.~Weinberg,
		``Effective gauge theories,''
		Phys. Lett. B \textbf{91}, 51-55 (1980).
%		doi:10.1016/0370-2693(80)90660-7
		%768 citations counted in INSPIRE as of 29 Aug 2025
		
		%\cite{Broncano:2002rw}
		\bibitem{Broncano:2002rw}
		A.~Broncano, M.~B.~Gavela and E.~E.~Jenkins,
		``The effective Lagrangian for the seesaw model of neutrino mass and leptogenesis,''
		Phys. Lett. B \textbf{552}, 177-184 (2003)
		[erratum: Phys. Lett. B \textbf{636}, 332 (2006)]
%		doi:10.1016/S0370-2693(02)03130-1
		[arXiv:hep-ph/0210271 [hep-ph]].
		%171 citations counted in INSPIRE as of 11 Sep 2025
		
		%\cite{Zhang:2021tsq}
		\bibitem{Zhang:2021tsq}
		D.~Zhang and S.~Zhou,
		``Radiative decays of charged leptons in the seesaw effective field theory with one-loop matching,''
		Phys. Lett. B \textbf{819}, 136463 (2021)
%		doi:10.1016/j.physletb.2021.136463
		[arXiv:2102.04954 [hep-ph]].
		%19 citations counted in INSPIRE as of 11 Jul 2025
		
		%\cite{Zhang:2021jdf}
		\bibitem{Zhang:2021jdf}
		D.~Zhang and S.~Zhou,
		``Complete one-loop matching of the type-I seesaw model onto the Standard Model effective field theory,''
		JHEP \textbf{09}, 163 (2021)
%		doi:10.1007/JHEP09(2021)163
		[arXiv:2107.12133 [hep-ph]].
		%52 citations counted in INSPIRE as of 30 Jul 2025
		
		%\cite{Wang:2023bdw}
		\bibitem{Wang:2023bdw}
		Y.~Wang, D.~Zhang and S.~Zhou,
		``Complete one-loop renormalization-group equations in the seesaw effective field theories,''
		JHEP \textbf{05}, 044 (2023)
%		doi:10.1007/JHEP05(2023)044
		[arXiv:2302.08140 [hep-ph]].
		%9 citations counted in INSPIRE as of 30 Jul 2025
		
		%\cite{Ibarra:2024tpt}
		\bibitem{Ibarra:2024tpt}
		A.~Ibarra, N.~Leister and D.~Zhang,
		``Complete two-loop renormalization group equation of the Weinberg operator,''
		JHEP \textbf{03}, 214 (2025)
%		doi:10.1007/JHEP03(2025)214
		[arXiv:2411.08011 [hep-ph]].
		%7 citations counted in INSPIRE as of 05 Aug 2025
		
		%\cite{Aoki:1982ed}
		\bibitem{Aoki:1982ed}
		K.~I.~Aoki, Z.~Hioki, R.~Kawabe, M.~Konuma and T.~Muta,
		``Electroweak Theory. Framework of On-Shell Renormalization and Study of Higher-Order Effects,''
		Prog. Theor. Phys. Suppl. \textbf{73}, 1-226 (1982).
%		doi:10.1143/PTPS.73.1
		%465 citations counted in INSPIRE as of 28 Aug 2025
		
		%\cite{Bohm:1986rj}
		\bibitem{Bohm:1986rj}
		M.~B\"{o}hm, H.~Spiesberger and W.~Hollik,
		``On the 1-Loop Renormalization of the Electroweak Standard Model and its Application to Leptonic Processes,''
		Fortsch. Phys. \textbf{34}, 687-751 (1986).
%		doi:10.1002/prop.19860341102
		%598 citations counted in INSPIRE as of 04 Sep 2025
		
		%\cite{Hollik:1988ii}
		\bibitem{Hollik:1988ii}
		W.~F.~L.~Hollik,
		``Radiative Corrections in the Standard Model and Their R\^{o}le for Precision Tests of the Electroweak Theory,''
		Fortsch. Phys. \textbf{38}, 165-260 (1990).
%		doi:10.1002/prop.2190380302
		%627 citations counted in INSPIRE as of 11 Aug 2025
		
		%\cite{Denner:1991kt}
		\bibitem{Denner:1991kt}
		A.~Denner,
		``Techniques for Calculation of Electroweak Radiative Corrections at the One-Loop Level and Results for $W$-physics at LEP 200,''
		Fortsch. Phys. \textbf{41}, 307-420 (1993)
%		doi:10.1002/prop.2190410402
		[arXiv:0709.1075 [hep-ph]].
		%1214 citations counted in INSPIRE as of 11 Sep 2025
		
		%\cite{Bohm:2001yx}
		\bibitem{Bohm:2001yx}
		M.~B\"{o}hm, A.~Denner and H.~Joos,
		``Gauge Theories of the Strong and Electroweak Interaction,''
		Vieweg+Teubner Verlag Wiesbaden, 2001.
%		doi:10.1007/978-3-322-80160-9
		%80 citations counted in INSPIRE as of 15 Aug 2025
		
		%\cite{Sirlin:2012mh}
		\bibitem{Sirlin:2012mh}
		A.~Sirlin and A.~Ferroglia,
		``Radiative corrections in precision electroweak physics: A historical perspective,''
		Rev. Mod. Phys. \textbf{85}, no.1, 263-297 (2013)
%		doi:10.1103/RevModPhys.85.263
		[arXiv:1210.5296 [hep-ph]].
		%76 citations counted in INSPIRE as of 25 Aug 2025
		
		%\cite{Denner:2019vbn}
		\bibitem{Denner:2019vbn}
		A.~Denner and S.~Dittmaier,
		``Electroweak radiative corrections for collider physics,''
		Phys. Rept. \textbf{864}, 1-163 (2020)
%		doi:10.1016/j.physrep.2020.04.001
		[arXiv:1912.06823 [hep-ph]].
		%233 citations counted in INSPIRE as of 29 Aug 2025
		
		%\cite{Huang:2023nqf}
		\bibitem{Huang:2023nqf}
		J.~Huang and S.~Zhou,
		``Mikheyev-Smirnov-Wolfenstein matter potential at the one-loop level in the Standard Model,''
		Phys. Rev. D \textbf{108}, no.9, 093010 (2023)
%		doi:10.1103/PhysRevD.108.093010
		[arXiv:2307.04685 [hep-ph]].
		%6 citations counted in INSPIRE as of 11 Sep 2025
		
		%\cite{Kniehl:1996bd}
		\bibitem{Kniehl:1996bd}
		B.~A.~Kniehl and A.~Pilaftsis,
		``Mixing renormalization in Majorana neutrino theories,''
		Nucl. Phys. B \textbf{474}, 286-308 (1996)
%		doi:10.1016/0550-3213(96)00280-5
		[arXiv:hep-ph/9601390 [hep-ph]].
		%111 citations counted in INSPIRE as of 15 Aug 2025
		
		%\cite{Pilaftsis:2002nc}
		\bibitem{Pilaftsis:2002nc}
		A.~Pilaftsis,
		``Gauge and scheme dependence of mixing matrix renormalization,''
		Phys. Rev. D \textbf{65}, 115013 (2002)
%		doi:10.1103/PhysRevD.65.115013
		[arXiv:hep-ph/0203210 [hep-ph]].
		%54 citations counted in INSPIRE as of 30 Jul 2025
		
		%\cite{Almasy:2009kn}
		\bibitem{Almasy:2009kn}
		A.~A.~Almasy, B.~A.~Kniehl and A.~Sirlin,
		``On-shell renormalization of the mixing matrices in Majorana neutrino theories,''
		Nucl. Phys. B \textbf{818}, 115-134 (2009)
%		doi:10.1016/j.nuclphysb.2009.03.025
		[arXiv:0902.3793 [hep-ph]].
		%14 citations counted in INSPIRE as of 11 Jul 2025
		
		%\cite{Martin:2019lqd}
		\bibitem{Martin:2019lqd}
		S.~P.~Martin and D.~G.~Robertson,
		``Standard model parameters in the tadpole-free pure $\overline{\rm{MS}}$ scheme,''
		Phys. Rev. D \textbf{100}, no.7, 073004 (2019)
%		doi:10.1103/PhysRevD.100.073004
		[arXiv:1907.02500 [hep-ph]].
		%36 citations counted in INSPIRE as of 04 Sep 2025
		
		%\cite{Alam:2022cdv}
		\bibitem{Alam:2022cdv}
		Z.~Alam and S.~P.~Martin,
		``Standard model at 200~GeV,''
		Phys. Rev. D \textbf{107}, no.1, 013010 (2023)
%		doi:10.1103/PhysRevD.107.013010
		[arXiv:2211.08576 [hep-ph]].
		%14 citations counted in INSPIRE as of 04 Sep 2025
		
		%\cite{Cabibbo:1963yz}
		\bibitem{Cabibbo:1963yz}
		N.~Cabibbo,
		``Unitary Symmetry and Leptonic Decays,''
		Phys. Rev. Lett. \textbf{10}, 531-533 (1963).
%		doi:10.1103/PhysRevLett.10.531
		%8037 citations counted in INSPIRE as of 12 Sep 2025
		
		%\cite{Kobayashi:1973fv}
		\bibitem{Kobayashi:1973fv}
		M.~Kobayashi and T.~Maskawa,
		``$CP$-Violation in the Renormalizable Theory of Weak Interaction,''
		Prog. Theor. Phys. \textbf{49}, 652-657 (1973).
%		doi:10.1143/PTP.49.652
		%12480 citations counted in INSPIRE as of 12 Sep 2025
		
		%\cite{Fleischer:1980ub}
		\bibitem{Fleischer:1980ub}
		J.~Fleischer and F.~Jegerlehner,
		``Radiative corrections to Higgs-boson decays in the Weinberg-Salam Model,''
		Phys. Rev. D \textbf{23}, 2001-2026 (1981).
%		doi:10.1103/PhysRevD.23.2001
		%379 citations counted in INSPIRE as of 04 Sep 2025
		
		%\cite{Bollini:1972ui}
		\bibitem{Bollini:1972ui}
		C.~G.~Bollini and J.~J.~Giambiagi,
		``Dimensional Renormalization: The Number of Dimensions as a Regularizing Parameter,''
		Nuovo Cim. B \textbf{12}, 20-26 (1972).
%		doi:10.1007/BF02895558
		%1142 citations counted in INSPIRE as of 04 Sep 2025
		
		%\cite{tHooft:1972tcz}
		\bibitem{tHooft:1972tcz}
		G.~'t Hooft and M.~J.~G.~Veltman,
		``Regularization and renormalization of gauge fields,''
		Nucl. Phys. B \textbf{44}, 189-213 (1972).
%		doi:10.1016/0550-3213(72)90279-9
		%5598 citations counted in INSPIRE as of 11 Sep 2025
		
		%\cite{Marciano:1980pb}
		\bibitem{Marciano:1980pb}
		W.~J.~Marciano and A.~Sirlin,
		``Radiative corrections to neutrino-induced neutral-current phenomena in the ${\rm SU}(2)_{L} \times {\rm U}(1)$ theory,''
		Phys. Rev. D \textbf{22}, 2695 (1980)
		[erratum: Phys. Rev. D \textbf{31}, 213 (1985)].
%		doi:10.1103/PhysRevD.22.2695
		%997 citations counted in INSPIRE as of 09 Sep 2025
		
		%\cite{Degrassi:1992ff}
		\bibitem{Degrassi:1992ff}
		G.~Degrassi and A.~Sirlin,
		``Gauge dependence of basic electroweak corrections of the Standard Model,''
		Nucl. Phys. B \textbf{383}, 73-92 (1992).
%		doi:10.1016/0550-3213(92)90671-W
		%97 citations counted in INSPIRE as of 30 Jul 2025
		
		%\cite{Denner:1990yz}
		\bibitem{Denner:1990yz}
		A.~Denner and T.~Sack,
		``Renormalization of the quark mixing matrix,''
		Nucl. Phys. B \textbf{347}, 203-216 (1990).
%		doi:10.1016/0550-3213(90)90557-T
		%132 citations counted in INSPIRE as of 12 Sep 2025
		
		%\cite{Pontecorvo:1957cp}
		\bibitem{Pontecorvo:1957cp}
		B.~Pontecorvo,
		``Mesonium and Antimesonium,''
		Sov. Phys. JETP \textbf{6}, 429-431 (1958).
		%2564 citations counted in INSPIRE as of 15 Sep 2025
		
		%\cite{Maki:1962mu}
		\bibitem{Maki:1962mu}
		Z.~Maki, M.~Nakagawa and S.~Sakata,
		``Remarks on the Unified Model of Elementary Particles,''
		Prog. Theor. Phys. \textbf{28}, 870-880 (1962).
%		doi:10.1143/PTP.28.870
		%5433 citations counted in INSPIRE as of 15 Sep 2025
		
		%\cite{Pontecorvo:1967fh}
		\bibitem{Pontecorvo:1967fh}
		B.~Pontecorvo,
		``Neutrino Experiments and the Problem of Conservation of Leptonic Charge,''
		Zh. Eksp. Teor. Fiz. \textbf{53}, 1717-1725 (1967).
		%2733 citations counted in INSPIRE as of 15 Sep 2025
		
		%\cite{Gambino:1998ec}
		\bibitem{Gambino:1998ec}
		P.~Gambino, P.~A.~Grassi and F.~Madricardo,
		``Fermion mixing renormalization and gauge invariance,''
		Phys. Lett. B \textbf{454}, 98-104 (1999)
%		doi:10.1016/S0370-2693(99)00321-4
		[arXiv:hep-ph/9811470 [hep-ph]].
		%98 citations counted in INSPIRE as of 11 Jul 2025
		
		%\cite{Kniehl:2000rb}
		\bibitem{Kniehl:2000rb}
		B.~A.~Kniehl, F.~Madricardo and M.~Steinhauser,
		``Gauge-independent $W$-boson partial decay widths,''
		Phys. Rev. D \textbf{62}, 073010 (2000)
%		doi:10.1103/PhysRevD.62.073010
		[arXiv:hep-ph/0005060 [hep-ph]].
		%66 citations counted in INSPIRE as of 11 Jul 2025
		
		%\cite{Yamada:2001px}
		\bibitem{Yamada:2001px}
		Y.~Yamada,
		``Gauge dependence of the on-shell renormalized mixing matrices,''
		Phys. Rev. D \textbf{64}, 036008 (2001)
%		doi:10.1103/PhysRevD.64.036008
		[arXiv:hep-ph/0103046 [hep-ph]].
		%95 citations counted in INSPIRE as of 11 Jul 2025
		
		%\cite{Liao:2003jy}
		\bibitem{Liao:2003jy}
		Y.~Liao,
		``Note on CKM matrix renormalization,''
		Phys. Rev. D \textbf{69}, 016001 (2004)
%		doi:10.1103/PhysRevD.69.016001
		[arXiv:hep-ph/0309034 [hep-ph]].
		%7 citations counted in INSPIRE as of 11 Jul 2025
		
		%\cite{Kniehl:2006bs}
		\bibitem{Kniehl:2006bs}
		B.~A.~Kniehl and A.~Sirlin,
		``Simple Approach to Renormalize the Cabibbo-Kobayashi-Maskawa Matrix,''
		Phys. Rev. Lett. \textbf{97}, 221801 (2006)
%		doi:10.1103/PhysRevLett.97.221801
		[arXiv:hep-ph/0608306 [hep-ph]].
		%20 citations counted in INSPIRE as of 11 Jul 2025
		
		%\cite{Kniehl:2006rc}
		\bibitem{Kniehl:2006rc}
		B.~A.~Kniehl and A.~Sirlin,
		``Simple on-shell renormalization framework for the Cabibbo-Kobayashi-Maskawa matrix,''
		Phys. Rev. D \textbf{74}, 116003 (2006)
%		doi:10.1103/PhysRevD.74.116003
		[arXiv:hep-th/0612033 [hep-th]].
		%35 citations counted in INSPIRE as of 11 Jul 2025
		
		%\cite{Almasy:2008ep}
		\bibitem{Almasy:2008ep}
		A.~A.~Almasy, B.~A.~Kniehl and A.~Sirlin,
		``Quark-mixing renormalization effects in the determination of the CKM parameters $|V_{ij}|$,''
		Phys. Rev. D \textbf{79}, 076007 (2009)
		[erratum: Phys. Rev. D \textbf{82}, 059901 (2010)]
%		doi:10.1103/PhysRevD.79.076007
		[arXiv:0811.0355 [hep-ph]].
		%12 citations counted in INSPIRE as of 11 Jul 2025
		
		%\cite{Kniehl:2009kk}
		\bibitem{Kniehl:2009kk}
		B.~A.~Kniehl and A.~Sirlin,
		``A novel formulation of Cabibbo-Kobayashi-Maskawa matrix renormalization,''
		Phys. Lett. B \textbf{673}, 208-210 (2009)
%		doi:10.1016/j.physletb.2009.02.024
		[arXiv:0901.0114 [hep-ph]].
		%25 citations counted in INSPIRE as of 11 Jul 2025
		
		%\cite{Almasy:2011vy}
		\bibitem{Almasy:2011vy}
		A.~A.~Almasy, B.~A.~Kniehl and A.~Sirlin,
		``Quark mixing renormalization effects in the determination of $|V_{tq}|$,''
		Phys. Rev. D \textbf{83}, 096004 (2011)
%		doi:10.1103/PhysRevD.83.096004
		[arXiv:1101.5758 [hep-ph]].
		%3 citations counted in INSPIRE as of 11 Jul 2025
		
		%\cite{Passarino:1978jh}
		\bibitem{Passarino:1978jh}
		G.~Passarino and M.~J.~G.~Veltman,
		``One-loop corrections for ${\rm e}^+ {\rm e}^-$ annihilation into $\mu^+ \mu^-$ in the Weinberg model,''
		Nucl. Phys. B \textbf{160}, 151-207 (1979).
%		doi:10.1016/0550-3213(79)90234-7
		%3089 citations counted in INSPIRE as of 12 Sep 2025
		
		%\cite{Sirlin:1991fd}
		\bibitem{Sirlin:1991fd}
		A.~Sirlin,
		``Theoretical Considerations Concerning the $Z^0$ Mass,''
		Phys. Rev. Lett. \textbf{67}, no.16, 2127-2130 (1991).
%		doi:10.1103/PhysRevLett.67.2127
		%253 citations counted in INSPIRE as of 16 Sep 2025
		
		%\cite{Stuart:1991xk}
		\bibitem{Stuart:1991xk}
		R.~G.~Stuart,
		``Gauge invariance, analyticity and physical observables at the ${\rm Z}^0$ resonance,''
		Phys. Lett. B \textbf{262}, 113-119 (1991).
%		doi:10.1016/0370-2693(91)90653-8
		%376 citations counted in INSPIRE as of 16 Sep 2025
		
		%\cite{Sirlin:1991rt}
		\bibitem{Sirlin:1991rt}
		A.~Sirlin,
		``Observations concerning mass renormalization in the electroweak theory,''
		Phys. Lett. B \textbf{267}, 240-242 (1991).
%		doi:10.1016/0370-2693(91)91254-S
		%133 citations counted in INSPIRE as of 16 Sep 2025
				
		%\cite{ParticleDataGroup:2024cfk}
		\bibitem{ParticleDataGroup:2024cfk}
		S.~Navas \textit{et al.} [Particle Data Group],
		``Review of particle physics,''
		Phys. Rev. D \textbf{110}, no.3, 030001 (2024).
%		doi:10.1103/PhysRevD.110.030001
		%2577 citations counted in INSPIRE as of 15 Sep 2025
		
		%\cite{Chetyrkin:2000yt}
		\bibitem{Chetyrkin:2000yt}
		K.~G.~Chetyrkin, J.~H.~K\"{u}hn and M.~Steinhauser,
		``{\tt RunDec}: a {\tt Mathematica} package for running and decoupling of the strong coupling and quark masses,''
		Comput. Phys. Commun. \textbf{133}, 43-65 (2000)
%		doi:10.1016/S0010-4655(00)00155-7
		[arXiv:hep-ph/0004189 [hep-ph]].
		%708 citations counted in INSPIRE as of 05 Sep 2025
		
		%\cite{Xing:2007fb}
		\bibitem{Xing:2007fb}
		Z.~z.~Xing, H.~Zhang and S.~Zhou,
		``Updated values of running quark and lepton masses,''
		Phys. Rev. D \textbf{77}, 113016 (2008)
%		doi:10.1103/PhysRevD.77.113016
		[arXiv:0712.1419 [hep-ph]].
		%445 citations counted in INSPIRE as of 08 Aug 2025
		
		%\cite{Xing:2011aa}
		\bibitem{Xing:2011aa}
		Z.~z.~Xing, H.~Zhang and S.~Zhou,
		``Impacts of the Higgs mass on vacuum stability, running fermion masses, and two-body Higgs decays,''
		Phys. Rev. D \textbf{86}, 013013 (2012)
%		doi:10.1103/PhysRevD.86.013013
		[arXiv:1112.3112 [hep-ph]].
		%172 citations counted in INSPIRE as of 27 Aug 2025
		
		%\cite{Herren:2017osy}
		\bibitem{Herren:2017osy}
		F.~Herren and M.~Steinhauser,
		``Version 3 of {\tt RunDec} and {\tt CRunDec},''
		Comput. Phys. Commun. \textbf{224}, 333-345 (2018)
%		doi:10.1016/j.cpc.2017.11.014
		[arXiv:1703.03751 [hep-ph]].
		%291 citations counted in INSPIRE as of 09 Sep 2025
		
		%\cite{Huang:2020hdv}
		\bibitem{Huang:2020hdv}
		G.~y.~Huang and S.~Zhou,
		``Precise values of running quark and lepton masses in the standard model,''
		Phys. Rev. D \textbf{103}, no.1, 016010 (2021)
%		doi:10.1103/PhysRevD.103.016010
		[arXiv:2009.04851 [hep-ph]].
		%70 citations counted in INSPIRE as of 04 Sep 2025
		
		%\cite{Fanchiotti:1992tu}
		\bibitem{Fanchiotti:1992tu}
		S.~Fanchiotti, B.~A.~Kniehl and A.~Sirlin,
		``Incorporation of QCD effects in basic corrections of the electroweak theory,''
		Phys. Rev. D \textbf{48}, 307-331 (1993)
%		doi:10.1103/PhysRevD.48.307
		[arXiv:hep-ph/9212285 [hep-ph]].
		%207 citations counted in INSPIRE as of 18 Aug 2025
		
		%\cite{Erler:1998sy}
		\bibitem{Erler:1998sy}
		J.~Erler,
		``Calculation of the QED coupling $\hat{\alpha} (M_Z)$ in the modified minimal-subtraction scheme,''
		Phys. Rev. D \textbf{59}, 054008 (1999)
%		doi:10.1103/PhysRevD.59.054008
		[arXiv:hep-ph/9803453 [hep-ph]].
		%172 citations counted in INSPIRE as of 11 Jul 2025
		
		%\cite{Degrassi:2003rw}
		\bibitem{Degrassi:2003rw}
		G.~Degrassi and A.~Vicini,
		``Two-loop renormalization of the electric charge in the standard model,''
		Phys. Rev. D \textbf{69}, 073007 (2004)
%		doi:10.1103/PhysRevD.69.073007
		[arXiv:hep-ph/0307122 [hep-ph]].
		%44 citations counted in INSPIRE as of 11 Jul 2025
		
		%\cite{Degrassi:2014sxa}
		\bibitem{Degrassi:2014sxa}
		G.~Degrassi, P.~Gambino and P.~P.~Giardino,
		``The $m_{W}-m_{Z}$ interdependence in the Standard Model: a new scrutiny,''
		JHEP \textbf{05}, 154 (2015)
%		doi:10.1007/JHEP05(2015)154
		[arXiv:1411.7040 [hep-ph]].
		%59 citations counted in INSPIRE as of 11 Aug 2025
		
		%\cite{Martin:2018yow}
		\bibitem{Martin:2018yow}
		S.~P.~Martin,
		``Matching relations for decoupling in the standard model at two loops and beyond,''
		Phys. Rev. D \textbf{99}, no.3, 033007 (2019)
%		doi:10.1103/PhysRevD.99.033007
		[arXiv:1812.04100 [hep-ph]].
		%12 citations counted in INSPIRE as of 04 Sep 2025
		
		%\cite{Jegerlehner:2001fb}
		\bibitem{Jegerlehner:2001fb}
		F.~Jegerlehner, M.~Y.~Kalmykov and O.~Veretin,
		``$\overline{\rm{MS}}$ vs. pole masses of gauge bosons: electroweak bosonic two-loop corrections,''
		Nucl. Phys. B \textbf{641}, 285-326 (2002)
%		doi:10.1016/S0550-3213(02)00613-2
		[arXiv:hep-ph/0105304 [hep-ph]].
		%124 citations counted in INSPIRE as of 04 Sep 2025
		
		%\cite{Jegerlehner:2002em}
		\bibitem{Jegerlehner:2002em}
		F.~Jegerlehner, M.~Y.~Kalmykov and O.~Veretin,
		``$\overline{\rm{MS}}$ vs. pole masses of gauge bosons II: two-loop electroweak fermion corrections,''
		Nucl. Phys. B \textbf{658}, 49-112 (2003)
%		doi:10.1016/S0550-3213(03)00177-9
		[arXiv:hep-ph/0212319 [hep-ph]].
		%155 citations counted in INSPIRE as of 04 Sep 2025
		
		%\cite{Kniehl:2015nwa}
		\bibitem{Kniehl:2015nwa}
		B.~A.~Kniehl, A.~F.~Pikelner and O.~L.~Veretin,
		``Two-loop electroweak threshold corrections in the Standard Model,''
		Nucl. Phys. B \textbf{896}, 19-51 (2015)
%		doi:10.1016/j.nuclphysb.2015.04.010
		[arXiv:1503.02138 [hep-ph]].
		%61 citations counted in INSPIRE as of 04 Sep 2025
		
		%\cite{Martin:2015lxa}
		\bibitem{Martin:2015lxa}
		S.~P.~Martin,
		``Pole mass of the $W$ boson at two-loop order in the pure $\overline{\rm MS}$ scheme,''
		Phys. Rev. D \textbf{91}, no.11, 114003 (2015)
%		doi:10.1103/PhysRevD.91.114003
		[arXiv:1503.03782 [hep-ph]].
		%28 citations counted in INSPIRE as of 04 Sep 2025
		
		%\cite{Martin:2015rea}
		\bibitem{Martin:2015rea}
		S.~P.~Martin,
		``$Z$-boson pole mass at two-loop order in the pure $\overline{\rm MS}$ scheme,''
		Phys. Rev. D \textbf{92}, no.1, 014026 (2015)
%		doi:10.1103/PhysRevD.92.014026
		[arXiv:1505.04833 [hep-ph]].
		%29 citations counted in INSPIRE as of 04 Sep 2025
		
		%\cite{Casas:1994qy}
		\bibitem{Casas:1994qy}
		J.~A.~Casas, J.~R.~Espinosa and M.~Quir\'{o}s,
		``Improved Higgs mass stability bound in the Standard Model and implications for Supersymmetry,''
		Phys. Lett. B \textbf{342}, 171-179 (1995)
%		doi:10.1016/0370-2693(94)01404-Z
		[arXiv:hep-ph/9409458 [hep-ph]].
		%524 citations counted in INSPIRE as of 25 Jul 2025
		
		%\cite{Bezrukov:2012sa}
		\bibitem{Bezrukov:2012sa}
		F.~Bezrukov, M.~Y.~Kalmykov, B.~A.~Kniehl and M.~Shaposhnikov,
		``Higgs boson mass and new physics,''
		JHEP \textbf{10}, 140 (2012)
%		doi:10.1007/JHEP10(2012)140
		[arXiv:1205.2893 [hep-ph]].
		%703 citations counted in INSPIRE as of 09 Sep 2025
		
		%\cite{Degrassi:2012ry}
		\bibitem{Degrassi:2012ry}
		G.~Degrassi, S.~Di Vita, J.~Elias-Mir\'{o}, J.~R.~Espinosa, G.~F.~Giudice, G.~Isidori and A.~Strumia,
		``Higgs mass and vacuum stability in the Standard Model at NNLO,''
		JHEP \textbf{08}, 098 (2012)
%		doi:10.1007/JHEP08(2012)098
		[arXiv:1205.6497 [hep-ph]].
		%1826 citations counted in INSPIRE as of 12 Sep 2025
		
		%\cite{Buttazzo:2013uya}
		\bibitem{Buttazzo:2013uya}
		D.~Buttazzo, G.~Degrassi, P.~P.~Giardino, G.~F.~Giudice, F.~Sala, A.~Salvio and A.~Strumia,
		``Investigating the near-criticality of the Higgs boson,''
		JHEP \textbf{12}, 089 (2013)
%		doi:10.1007/JHEP12(2013)089
		[arXiv:1307.3536 [hep-ph]].
		%1520 citations counted in INSPIRE as of 12 Sep 2025
		
		%\cite{Martin:2014cxa}
		\bibitem{Martin:2014cxa}
		S.~P.~Martin and D.~G.~Robertson,
		``Higgs boson mass in the standard model at two-loop order and beyond,''
		Phys. Rev. D \textbf{90}, no.7, 073010 (2014)
%		doi:10.1103/PhysRevD.90.073010
		[arXiv:1407.4336 [hep-ph]].
		%60 citations counted in INSPIRE as of 04 Sep 2025
		
		%\cite{Hempfling:1994ar}
		\bibitem{Hempfling:1994ar}
		R.~Hempfling and B.~A.~Kniehl,
		``Relation between the fermion pole mass and $\overline{\rm MS}$ Yukawa coupling in the standard model,''
		Phys. Rev. D \textbf{51}, no.3, 1386-1394 (1995)
%		doi:10.1103/PhysRevD.51.1386
		[arXiv:hep-ph/9408313 [hep-ph]].
		%147 citations counted in INSPIRE as of 11 Jul 2025
		
		%\cite{Jegerlehner:2003py}
		\bibitem{Jegerlehner:2003py}
		F.~Jegerlehner and M.~Y.~Kalmykov,
		``The $O(\alpha \alpha_s)$ correction to the pole mass of the $t$-quark within the Standard Model,''
		Nucl. Phys. B \textbf{676}, 365-389 (2004)
%		doi:10.1016/j.nuclphysb.2003.10.012
		[arXiv:hep-ph/0308216 [hep-ph]].
		%104 citations counted in INSPIRE as of 11 Jul 2025
		
		%\cite{Jegerlehner:2003sp}
		\bibitem{Jegerlehner:2003sp}
		F.~Jegerlehner and M.~Y.~Kalmykov,
		``$O(\alpha \alpha_{\rm s})$ Relation Between Pole- and $\overline{\rm{MS}}$-Mass of the $t$-Quark,''
		Acta Phys. Polon. B \textbf{34}, 5335-5344 (2003)
		[arXiv:hep-ph/0310361 [hep-ph]].
		%39 citations counted in INSPIRE as of 11 Jul 2025
		
		%\cite{Eiras:2005yt}
		\bibitem{Eiras:2005yt}
		D.~Eiras and M.~Steinhauser,
		``Two-loop ${\cal O}(\alpha \alpha_s)$ corrections to the on-shell fermion propagator in the standard model,''
		JHEP \textbf{02}, 010 (2006)
%		doi:10.1088/1126-6708/2006/02/010
		[arXiv:hep-ph/0512099 [hep-ph]].
		%24 citations counted in INSPIRE as of 11 Jul 2025
		
		%\cite{Jegerlehner:2012kn}
		\bibitem{Jegerlehner:2012kn}
		F.~Jegerlehner, M.~Y.~Kalmykov and B.~A.~Kniehl,
		``On the difference between the pole and the $\overline{\rm MS}$ masses of the top quark at the electroweak scale,''
		Phys. Lett. B \textbf{722}, 123-129 (2013)
%		doi:10.1016/j.physletb.2013.04.012
		[arXiv:1212.4319 [hep-ph]].
		%86 citations counted in INSPIRE as of 19 Aug 2025
		
		%\cite{Faisst:2003px}
		\bibitem{Faisst:2003px}
		M.~Faisst, J.~H.~K\"{u}hn, T.~Seidensticker and O.~Veretin,
		``Three loop top quark contributions to the $\rho$ parameter,''
		Nucl. Phys. B \textbf{665}, 649-662 (2003)
%		doi:10.1016/S0550-3213(03)00450-4
		[arXiv:hep-ph/0302275 [hep-ph]].
		%162 citations counted in INSPIRE as of 23 Jul 2025
		
		%\cite{Faisst:2004gn}
		\bibitem{Faisst:2004gn}
		M.~Faisst, J.~H.~K\"{u}hn and O.~Veretin,
		``Pole- versus MS-mass definitions in the electroweak theory,''
		Phys. Lett. B \textbf{589}, 35-38 (2004)
%		doi:10.1016/j.physletb.2004.03.045
		[arXiv:hep-ph/0403026 [hep-ph]].
		%27 citations counted in INSPIRE as of 11 Jul 2025
		
		%\cite{Kataev:2022dua}
		\bibitem{Kataev:2022dua}
		A.~L.~Kataev and V.~S.~Molokoedov,
		``Notes on Interplay between the QCD and EW Perturbative Corrections to the Pole-Running-to-Top-Quark Mass Ratio,''
		JETP Lett. \textbf{115}, no.12, 704-712 (2022)
%		doi:10.1134/S0021364022600902
		[arXiv:2201.12073 [hep-ph]].
		%11 citations counted in INSPIRE as of 18 Sep 2025
		
		%\cite{Kniehl:2014yia}
		\bibitem{Kniehl:2014yia}
		B.~A.~Kniehl and O.~L.~Veretin,
		``Two-loop electroweak threshold corrections to the bottom and top Yukawa couplings,''
		Nucl. Phys. B \textbf{885}, 459-480 (2014)
		[erratum: Nucl. Phys. B \textbf{894}, 56-57 (2015)]
%		doi:10.1016/j.nuclphysb.2015.02.012
		[arXiv:1401.1844 [hep-ph]].
		%28 citations counted in INSPIRE as of 04 Sep 2025
		
		%\cite{Martin:2016xsp}
		\bibitem{Martin:2016xsp}
		S.~P.~Martin,
		``Top-quark pole mass in the tadpole-free $\overline{\rm MS}$ scheme,''
		Phys. Rev. D \textbf{93}, no.9, 094017 (2016)
%		doi:10.1103/PhysRevD.93.094017
		[arXiv:1604.01134 [hep-ph]].
		%30 citations counted in INSPIRE as of 04 Sep 2025
		
		%\cite{Arason:1991ic}
		\bibitem{Arason:1991ic}
		H.~Arason, D.~J.~Casta\~{n}o, B.~Keszthelyi, S.~Mikaelian, E.~J.~Piard, P.~Ramond and B.~D.~Wright,
		``Renormalization-group study of the standard model and its extensions: The standard model,''
		Phys. Rev. D \textbf{46}, no.9, 3945-3965 (1992).
%		doi:10.1103/PhysRevD.46.3945
		%471 citations counted in INSPIRE as of 04 Sep 2025
		
		%\cite{Gray:1990yh}
		\bibitem{Gray:1990yh}
		N.~Gray, D.~J.~Broadhurst, W.~Grafe and K.~Schilcher,
		``Three-loop relation of quark $\overline{\rm{MS}}$ and pole masses,''
		Z. Phys. C \textbf{48}, 673-679 (1990).
%		doi:10.1007/BF01614703
		%724 citations counted in INSPIRE as of 18 Aug 2025
		
		%\cite{Bekavac:2007tk}
		\bibitem{Bekavac:2007tk}
		S.~Bekavac, D.~Seidel, M.~Steinhauser and A.~Grozin
		``Light quark mass effects in the on-shell renormalization constants,''
		JHEP \textbf{10}, 006 (2007)
%		doi:10.1088/1126-6708/2007/10/006
		[arXiv:0708.1729 [hep-ph]].
		%64 citations counted in INSPIRE as of 11 Jul 2025
		
		%\cite{Kniehl:2004hfa}
		\bibitem{Kniehl:2004hfa}
		B.~A.~Kniehl, J.~H.~Piclum and M.~Steinhauser,
		``Relation between bottom-quark $\overline{\rm{MS}}$ Yukawa coupling and pole mass,''
		Nucl. Phys. B \textbf{695}, 199-216 (2004)
%		doi:10.1016/j.nuclphysb.2004.06.036
		[arXiv:hep-ph/0406254 [hep-ph]].
		%18 citations counted in INSPIRE as of 11 Jul 2025
		
		%\cite{Martin:2005ch}
		\bibitem{Martin:2005ch}
		S.~P.~Martin,
		``Fermion self-energies and pole masses at two-loop order in a general renormalizable theory with massless gauge bosons,''
		Phys. Rev. D \textbf{72}, no.9, 096008 (2005)
%		doi:10.1103/PhysRevD.72.096008
		[arXiv:hep-ph/0509115 [hep-ph]].
		%56 citations counted in INSPIRE as of 11 Jul 2025
		
		%\cite{Patel:2015tea}
		\bibitem{Patel:2015tea}
		H.~H.~Patel,
		``{\it Package}-X: A {\it Mathematica} package for the analytic calculation of one-loop integrals,''
		Comput. Phys. Commun. \textbf{197}, 276-290 (2015)
%		doi:10.1016/j.cpc.2015.08.017
		[arXiv:1503.01469 [hep-ph]].
		%581 citations counted in INSPIRE as of 12 Sep 2025
		
		%\cite{Patel:2016fam}
		\bibitem{Patel:2016fam}
		H.~H.~Patel,
		``{\it Package}-X 2.0: A {\it Mathematica} package for the analytic calculation of one-loop integrals,''
		Comput. Phys. Commun. \textbf{218}, 66-70 (2017)
%		doi:10.1016/j.cpc.2017.04.015
		[arXiv:1612.00009 [hep-ph]].
		%315 citations counted in INSPIRE as of 12 Sep 2025
	\end{thebibliography}
\end{document}